\documentclass{aa}  

\usepackage{graphicx}
\usepackage{txfonts}
\usepackage{subcaption}
\usepackage{float}
\usepackage[colorlinks=true,linkcolor=blue,citecolor=blue,urlcolor=blue]{hyperref}

\begin{document}

   \title{Constraining r-process nucleosynthesis with multi-objective Galactic chemical evolution models}

   \author{M. Molero
          \inst{1,2}
          \and
          A. Arcones\inst{1,3,4}
          \and
          F. Montes\inst{5,6}
          \and
          C. J. Hansen\inst{7}
          }

   \institute{Institut f\"ur Kernphysik, Technische Universit\"at Darmstadt, Schlossgartenstr. 2, Darmstadt 64289, Germany\\
              \email{marta.molero@tu-darmstadt.de; marta.molero@inaf.it}
        \and
            INAF, Osservatorio Astronomico di Trieste, Via Tiepolo 11, I-34131 Trieste, Italy
        \and
            GSI Helmholtzzentrum f¨ur Schwerionenforschung GmbH, 64291 Darmstadt, Germany
        \and
            Max-Planck-Institut f¨ur Kernphysik, Saupfercheckweg 1, 69117 Heidelberg, Germany
        \and
            National Superconducting Cyclotron Laboratory, East Lansing, MI 48824, USA
        \and
            Joint Institute for Nuclear Astrophysics—CEE, Michigan State University, East Lansing, MI 48824, USA
        \and
            Goethe University Frankfurt, Institute for Applied Physics (IAP), Max-von-Laue-Str. 12, 60438, Frankfurt am Main, Germany
             }


 
  \abstract
   {The astrophysical site(s) of the rapid neutron-capture process (r-process) remain uncertain, with competing scenarios such as neutron star mergers and magneto-rotational supernovae, offering different predictions for the timing, frequency and yield for heavy elements pollution. Galactic chemical evolution models provide an important tool to constrain these properties by comparing model predictions to the observed abundance of neutron-capture elements across different metallicities.}
   {Our aim is to explore, in a systematic and data-driven manner, the range of astrophysical conditions under which the r-process enrichment can reproduce the observed trends of multiple neutron-capture elements in the Milky Way. Rather than assuming a fixed physical site, we adopt a flexible, parametric approach to assess whether a common set of r-process parameters can simultaneously explain the chemical evolution of several heavy elements of interest.}
   {We construct a grid of one-infall, homogeneous, Galactic chemical evolution model varying four key parameters: the Eu r-process yield per event, the rate of r-process-producing events, the delay time before enrichment, and the progenitor star mass range. For each of the $\sim 1.5 \times 10^5$ models, we compute the predicted [X/Fe] vs. [Fe/H] trends for several neutron-capture elements, obtained by scaling Eu yields with the solar r-process pattern, and evaluate the model performance using $\chi^2$ statistics. A multi-objective optimization approach based on Pareto front analysis is used to identify models that best reproduce the observed abundance trends across multiple neutron-capture elements simultaneously.}
   {The best models consistently favour short delay times ($\leq 30\ \rm Myr$), low-mass progenitors ($\sim 20-25\ \rm M_\odot$) and effective Eu injection rate of $\sim 2 \times 10^{-7}\ \rm M_\odot$ per event. Stars more massive than $\sim 80\ \rm M_\odot$ are too rare to account alone for the observed r-process enrichment and therefore they can not be the only source. While heavier neutron-capture elements (Ba, La, Ce) can be simultaneously reproduced within the Pareto-optimal set, the lighter ones (Sr, Y, Zr) present stronger conflicts with Eu, leading to systematically larger distances from the ideal solution. This tension reflects the fact that the solar r-process scaling relation, used to extend Eu yields to other elements, becomes progressively less valid toward lighter neutron-capture elements.}
   {Our results suggest that no single class of r-process events, under solar-scaled yields, can simultaneously account for the chemical evolution of both light and heavy neutron-capture elements. Instead, at least two distinct scaling components appear necessary: one corresponding to the main r-process, similar to the solar and r-process rich stars patterns, and another weak component with enhanced production of lighter r-process elements, similar to that observed in r-process poor stars.}

   \authorrunning{M. Molero, A. Arcones, F. Montes, C. J. Hansen}
   \titlerunning{Constraining r-process nucleosynthesis}

   \maketitle

\section{Introduction}

The origin of the heavy elements produced by neutron-capture processes remains one of the central open questions in astrophysics. Two main channels have been identified: the slow neutron-capture process (s-process), operating mainly in asymptotic giant branch (AGB) stars (e.g., \citealp{Gallino1998, Lugaro2003, Cristallo2015, Karakas2016, Busso2021}) and rotating massive stars (e.g., \citealp{pignatari2010, 2016MNRAS.456.1803F, Limongi2018}), and the rapid neutron-capture process (r-process), whose astrophysical site(s) are still debated (see e.g. \citealp{Cote2019, Cowan2021, Bandyopadhyay2025}). Additional contributions from the intermediate neutron-capture process (i-process) may further shape the observed abundance patterns of some elements, particularly at low metallicities (e.g., \citealp{Remple2024, Choplin2024}).

Among these, the r-process is responsible for roughly half of the isotopes heavier than iron. Recent multi-messenger observations have firmly established neutron-star mergers (NSMs) as viable r-process sites (e.g., \citealp{Pian2017, Hotokezaka2018, Watson2019}), but their delayed occurrence and low observed rates challenges their ability to reproduce the presence of r-process material in the earliest stars (\citealp{Matteucci2014, Saleem2025, Chen2025}). Alternative or additional sources, such as rare classes of core-collapse supernovae (CC-SNe; e.g. magneto-rotational supernovae, MR-SNe, and/or collapsars; \citealp{Nishimura2017, mosta2018, Siegel2019, Reichert2021, Reichert2023}) or magnetar giant flares (\citealp{Cehula2024, Patel2025b, Patel2025}) remain under investigation. Disentangling the relative contribution of these candidate sites is a key step toward a comprehensive understanding of r-process nucleosynthesis.

Observations of abundances of neutron-capture elements in low-metallicity stars provide insights into the astrophysical origin of the r-process (\citealp{Francois2007, Barbuy2011, Hansen2012, Bandyopadhyay2024, Alencastro2025, Lombardo2025}). The large star-to-star scatter observed in heavy element ratios at low-metallicities is an indication of the rarity of the r-process site(s). Moreover, it suggests that many of those stars might have been enriched by only one or a few r-process events, and can then be used to validate theoretical nucleosynthesis predictions (\citealp{Magg2020, Hansen2020, Yong2021}). At later stages of galactic evolution, when successive generations of stars have enriched the interstellar medium (ISM), the chemical environment becomes progressively more homogeneous. In this regime, the stochastic imprint of individual r-process events is averaged out, and the abundance trends of neutron-capture elements instead reflect the integrated contribution of different nucleosynthetic channels over time. These include not only the different potential r-process sources, each with their characteristic delay timescales, but also the s-process, whose contribution becomes increasingly significant at higher metallicity. As a consequence, the chemical pattern observed in any given star is no longer a direct imprint of a single enrichment event, but rather the results of the effects of many enrichment sources, shaped by the specific history of the host galaxy, including its star formation, gas flows with their specific chemical composition and stellar migration (see \citealp{Matteucci2021} for a review).

Because of these overlapping contributions and their metallicity and/or time dependence, homogenous chemical evolution models are required to disentangle the relative role of the different nucleosynthetic channels. If the focus is on the r-process, one usually relies primarily on Eu, being the only neutron-capture element with extensive stellar observations that is almost exclusively produced by the r-process ($\sim 95\%$ at solar according to \citealp{Prantzos2020}). Therefore, its abundance trends provide a direct tracer of r-process enrichment across time. The common approach to reproduce the observed [Eu/Fe] vs. [Fe/H] trend is to include at least two distinct r-process sources: a delayed source, typically associated with NSMs, and a prompt source, such as MR-SNe. For the latter, the low expected frequency of events usually leads chemical evolution models to assume that only a small fraction ($\sim 10\% - 20\%$) of massive stars within a given mass range (e.g., $10-80\ \rm M_\odot$) end their lives as MR-SNe (e.g., \citealp{Cescutti2015, Kobayashi2023, Molero2023}). These quantities, namely the event fraction, the progenitor mass interval, and the Eu yield as well, are free quantities in the models, highly degenerate with one another, and unconstrained by direct observations (while nucleosynthesis calculations can in principle provide yield estimates, current simulations span a wide range of outcomes). In the case of NSMs, estimations of their yields and frequency can rely on some observations. The event rate can be constrained to reproduce the latest estimation from \citet{Abbott2021}, while the yields of r-process for Eu and other neutron-capture elements can be obtained by scaling to solar the abundance of Sr derived from the re-analysis of the spectra of the kilonova \textit{AT}2017gfo by \citet{Watson2019} (\citealp{Molero2021}). However, these observational constraints are still affected by significant uncertainties. The merger rate reported by \citet{Abbott2021} is measured as a cosmic rate and must be converted into a Galactic rate (see \citealp{Simonetti2019} for details on the conversion), and, most importantly, the current estimate relies on a very limited sample of detected events, which results in large statistical uncertainties. Moreover, fixing the Eu yield (and eventually that for the r-process component of other neutron-capture elements) by scaling the measured Sr one to solar r-process abundances, implicitly assumes that the concept of r-process universality can be extended to the lighter elements, and not restricted solely to the lanthanides and third-peak nuclei.

Observations of low-metallicity r-process rich stars (r- rich stars) such as CS 22892-052 \citep{Sneden2003} and HD 222925 \citep{Roederer2018}, show a strong agreement between the abundances of the heaviest neutron-capture elements (Ba and above, $Z \geq 56$) and the scaled solar r-process distribution, while the lighter neutron-capture elements show deviations from the solar pattern. This agreement between old stars and the solar distribution supports the robustness of the r-process over time for the elements heavier than Ba, which may arise from the fact that a wide range of astrophysical conditions can yield a similar abundance pattern \citep{Kuske2025}. In contrast, r-process poor stars (r- poor stars) such as HD 122563 \citep{Honda2006} and HD 88609 \citep{Honda2007}, deviate from the solar distribution, mostly because of an enhancement of the lighter neutron-capture elements when normalized to Eu. Overall, these observations indicate that the universality of the r-process does not extend across the entire mass range: the abundances of the heavier elements ($Z > 50$) appear more robust, while those of the lighter ones ($38 < Z < 50$) show considerable variation \citep[e.g.,][]{Hansen2014}. Analogously to the s-process, the r-process is divided into a weak component, primarily responsible for the production of the lighter neutron-capture elements, and a main component, which synthesizes the heavier elements (see \citealp{ArconesThielemann2023} for a review).

Testing such a scenario in a chemical evolution context, extending the analysis toward higher metallicity, requires moving beyond a single observational constraint. The observed abundance ratios trend of multiple neutron-capture elements can indeed provide complementary informations, with the heaviest being potentially predominantly tracers of the main r-process, while the lighter being more sensitive to the weak r-process. Given the large uncertainties associated with the possible astrophysical sites of the r-process we discussed above, in this work we do not assume one or more fixed physical sources. Instead, we adopt a flexible, parametric framework. Starting from a well-constrained, homogeneous Galactic chemical evolution model, we construct a grid of models that systematically varies the key parameters of the r-process production: the Eu yield per event, the event rate, the delay time for enrichment of the ISM, and the progenitor star mass range. The yields of the other neutron-capture elements are then obtained by scaling the best-Eu models to the solar r-process pattern. By confronting these models with observations of both heavy and light neutron-capture elements, we can directly test to what extent the solar-scaled relation remains valid across different metallicities and nucleosynthetic regimes. In this context, we treat the comparison between models and data as a multi-objective optimization problem, where each element provides an independent observational constraint. Rather than identifying a single best-fit model, the goal is to map the set of Pareto-optimal solutions, namely models for which improving the agreement with one element would necessarily worsen the agreement with another. In such a systematic way, we can explore the trade-offs among different observational constraints and identify the regions of parameter space most consistent with the different neutron-capture abundance ratio patterns.

The paper is organized as follows. In Section \ref{sec: observational data}, we present the observational datasets adopted for comparison with our models. Section \ref{sec: The chemical evolution model} describes the chemical evolution model and the parametrization of r-process nucleosynthesis. In section \ref{sec: Eu optimization}, we describe the optimization procedure and discuss the results of the best-Eu models. Section \ref{sec: Multi-objective optimization} presents the multi-objective optimization strategy and the Pareto formalism, while section \ref{sec: results} presents the corresponding results for the different elements. In section \ref{sec: different scaling relations}, we examine the impact of adopting alternative r-process patterns. Finally, section \ref{sec: conclusions} summarizes our main findings and conclusions.

\section{Observational data}
\label{sec: observational data}

For all neutron-capture elements, including Eu, there is currently no single survey that provides abundance measurements at high resolution across the full metallicity range in any given galactic environment. As a consequence, it is necessary to construct a reliable dataset by combining observations from multiple surveys, allowing us to sample the entire metallicity range of interest. It is important to note that, while combining data from different surveys allows us to cover a broader metallicity range, it also introduces certain limitations. In particular, the resulting dataset is unavoidably inhomogeneous, as the abundances may have been derived using different methods, line lists, model atmospheres, and overall assumptions, and the observational uncertainties may vary across studies. However, until a single survey spanning the full metallicity range becomes available, such a composite approach remains necessary, provided that data are carefully selected to ensure consistency as much as possible. This limitation was highlighted in \cite{Palla2025}. In this work, we adopt a similar compilation, with only minor additions, that now we will briefly describe.

In the intermediate-to-high metallicity range ($\mathrm{[Fe/H] \geq -1.25\ dex}$), our sample includes both field stars and open clusters (OCs) from the Gaia-ESO Survey \citep{Randich2022, Gilmore2022}. In particular, we adopt the OC sample used by \citet{Magrini2023}, which consists of 62 clusters older than 100 Myr, and the field star sample selected following the criteria of \citet{Viscasillas2022}. All selected Gaia-ESO stars lie within a Galactocentric radius range of $7 < R_{\rm GC}/\mathrm{kpc} < 9$. In a few cases where the Gaia-ESO survey does not provide measurements for specific elements of interest, stars from \citet{BattistiniBensby2016} are used instead.

The lower-metallicity regime ($\mathrm{-2.7\leq[Fe/H]<- 1.25}$) is populated using stars from the Measuring at Intermediate metallicity Neutron-Capture Elements (MINCE) survey (\citealp{Cescutti2022}), specifically from its second and third releases (MINCE II and III respectively, \citealp{Francois2024, Lucertini2025}). This is complemented with a collection of stars from individual studies selected from the Stellar Abundances for Galactic Archaeology (SAGA) database (\citealp{Suda2008}), including \cite{Mishenina2001, Hansen2012, Ishigaki2013, Roederer2014, Li2022}. Those were selected mainly by their resolution, which we imposed to be $R \geq 40\ 000$, to ensure consistency with the high values of the Gaia-ESO ($R \simeq 47\ 000$) and MINCE (up to $R \geq 100\ 000$) surveys. To reduce contamination from accreted stellar populations, we restricted our SAGA sample to stars likely formed in situ. This was achieved by integrating stellar orbits using the Galactic potential of \citet{McMillan2017} and applying the kinematic selection criteria described in \citet{Monty2024}.

The very low-metallicity regime ($\mathrm{[Fe/H] < -2.7\ dex}$) is not considered in this work. As discussed in the following section, we adopt a homogeneous chemical evolution model, which cannot capture the stochastic nature of early Galactic enrichment. The large scatter observed in neutron-capture element abundances at these low metallicities is believed to result from a combination of factors: the intrinsically inhomogeneous chemical evolution at early times and the potential rarity (and eventually spatial sparsity) of r-process events during the initial phases of Galactic history \citep{Cescutti2015}.

A further limitation comes to the fact that, while it is often possible to construct a reasonably complete dataset for Eu, similar coverage is rarely available for many other neutron-capture elements. As a result, although Eu can be used to constrain r-process enrichment, it becomes difficult to assess whether the same nucleosynthetic assumptions also reproduce the observed behaviour of other neutron-capture elements. In particular, while large datasets can be built for $1^{\rm st}$- and $2^{\rm nd}$-peak s-process elements (and for a few other cases), data for r-process peak elements remain few (or null). In most cases, the number of high-quality measurements is too low to allow a statistically significant comparison with chemical evolution models.

\section{The chemical evolution model}
\label{sec: The chemical evolution model}

In this work, since we aim to perform a parametric exploration of nucleosynthetic inputs, we adopt a one-infall Galactic chemical evolution model (see also \citealp{Prantzos2018}). However, if the interest is in more complex, multi-infall chemical evolution scenarios applied to neutron-capture elements, we refer to our previous studies (\citealp{Molero2023, Molero2025}) where such models are discussed in details (see also \citealp{Chiappini1997, Colavitti2008, Spitoni2021, Spitoni2023, Palla2024} for descriptions of most complex infall scenarios).

As the main purpose is to study the nucleosynthesis of neutron-capture elements, a one-infall chemical evolution model has several advantages over more complex multi-infall scenarios. In fact, the one-infall model provides a simplified and controlled evolutionary framework, which might be better suited for parametric studies where the aim is to systematically explore the impact of nucleosynthetic parameters (e.g., event rates, delay times, and yields) on the predicted abundance trends. A one-infall scenario minimizes the number of free parameters related to Galactic evolution (e.g., gas accretion history and star formation), reducing degeneracies and, in particular, allowing a more direct mapping between the model inputs and the observed abundance ratios trends. Also, a multiple infall scenario alters the metallicity evolution of the ISM, diluting the metal content. This has a strong impact on the s-process nucleosynthesis, highly sensitive to the initial stellar metallicity. In such scenarios, changes in abundance patterns can be driven not only by the nucleosynthesis itself, but also by the timing and efficiency of the infall events, complicating the interpretation of the results. As a consequence, although multi-infall models are more realistic in capturing the chemodynamical history of the Galaxy (but see \citealp{Dubay2025}), they introduce another layer of complexity and, by contrast, a one-infall model might ensure a smoother chemical evolution, making it easier to disentangle the contribution of different r- and s-process sources across time and metallicity.

The one-infall model (see e.g., \citealp{MatteucciFrancois1989}) assumes that the solar neighbourhood formed through a single episode of gas accretion with primordial composition, occurring on a timescale of $\tau_{\rm inf} \simeq 7\ \mathrm{Gyr}$. Stars of different masses form according to the initial mass function (IMF) of \citet{Kroupa1993}, and star formation proceeds following a Schmidt–Kennicutt law \citep{Kennicutt1998} of the form:
\begin{equation}
\psi(t) \propto \sigma_{\mathrm{gas}}(t)^k,
\end{equation}
where $\sigma_{\mathrm{gas}}(t)$ is the gas surface density, and the law-index is set to $k = 1.5$. The evolution of the time derivative of the surface mass density, $\dot{\sigma}_i(t)$, of a chemical element $i$ in the ISM is described by the following equation:
\begin{equation}
\dot{\sigma}_i(t) = -\psi(t) X_i(t) + X_{\mathrm{inf},i} A(t) + R_i(t),
\end{equation}
where $\sigma_i(t)$ is the surface density of the element $i$, and $X_i(t)$ is the mass fraction of the element $i$ at time $t$. The term $X_{\mathrm{inf},i}$ represents the mass fraction of the element $i$ in the infalling gas, which, as stated above, is assumed to be of primordial composition. The gas accretion rate is described by the function $A(t)$, which, in the one-infall model, follows a simple exponentially decreasing law:
\begin{equation}
    A(t)=ae^{-\frac{t}{\tau_{\rm inf}}},
\end{equation}
where $\tau_{\rm inf}$ is the infall timescale, and the normalization constant $a$ is chosen to reproduce the present-day surface mass density of the Galactic disk in the solar vicinity. Finally, $R_i(t)$ is the rate at which the element $i$ is returned to the ISM by dying stars. It accounts for the contributions from various nucleosynthetic phenomena, including Type Ia and CC-SNe, stellar winds, novae and, in general, MR-SNe and NSMs. A detailed formulation of this term can be found in \citet{Matteucci2012}. As described in the following section, in this work we relax the assumption that the r-process originates from a specific astrophysical site. Instead, we adopt a more flexible, parametric approach to model its contribution. As a consequence, the inclusion of MR-SNe and NSMs does not follow the classical site-based prescription, for which we refer instead to \citet{Molero2023}.

Concerning the nucleosynthesis prescriptions from the other sites, here we adopt:
\begin{itemize}
    \item Yields from the non-rotational set of the FRUITY database (\citealp{Cristallo2009, Cristallo2011, Cristallo2015}) for low- and intermediate-mass stars (LIMSs; with initial masses $\mathrm{1.3\leq M/M_\odot\leq6.0}$; see section \ref{sec: s-process nucleosynthesis}).
    \item \citet{Doherty2014} yields for super-AGB stars between $\mathrm{6.5\leq M/M_\odot\leq9.0}$
    \item For massive stars ($\mathrm{13 \leq M/M_\odot \leq 100}$), we adopt the rotating yields from "set R" of \citet{Limongi2018}. The initial rotational velocities are distributed according to the metallicity-dependent prescription of \citet{Prantzos2018}, which assumes that massive stars rotate faster towards lower metallicities (see also \citealp{Romano2019, Rizzuti2021, Molero2024}; see section \ref{sec: s-process nucleosynthesis}).
    \item For Type Ia SNe, we assume the single-degenerate scenario for the progenitors, with stellar yields from \citet{Iwamoto1999}.
    \item For novae, which do not affect the nucleosynthesis of the heavy elements we are interested here, we adopt yields from \citet{Jose2007}.
\end{itemize}

\subsection{S-process nucleosynthesis}
\label{sec: s-process nucleosynthesis}

\begin{figure*}
    \includegraphics[width=1\textwidth]{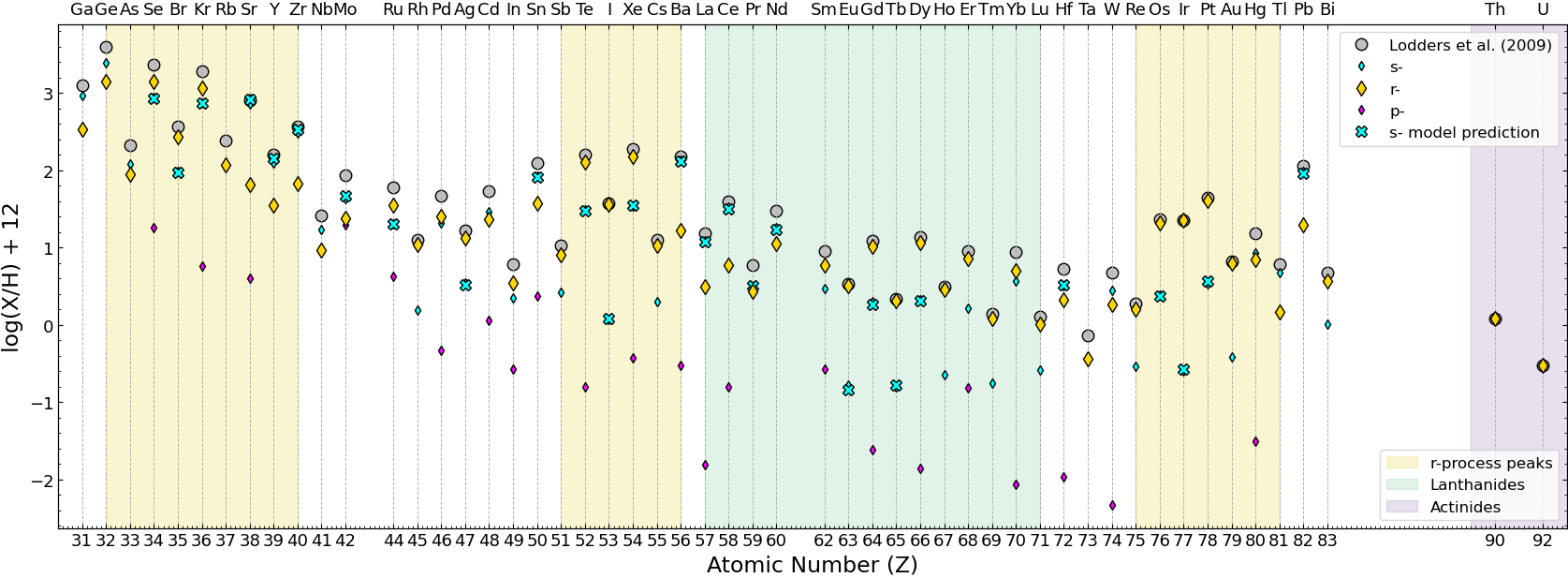}
     \caption{\citet{Lodders2009} total solar abundances of neutron-capture elements (grey dots) with the corresponding r-, s- and p-process fractions (yellow, cyan and magenta diamonds, respectively) from \citet{Prantzos2020}. Also shown are the chemical evolution model predictions for the s-process component for the subset of elements included in this work (cyan crosses).}%
 \label{fig: solar fraction}%
\end{figure*}

Since the focus of this work is on the r-process, the s-process nucleosynthetic prescriptions are kept fixed across all models. Unlike the r-process, the s-process benefits from extensive grids of stellar models covering a wide range of progenitor masses and metallicities. This significantly reduces the need for arbitrary interpolations and limits the associated uncertainties in our model predictions. The two main sources of s-process nucleosynthesis in our model are LIMS during the AGB phase, and rotating massive stars. The yield prescriptions, as described above, are based on model grids that both cover $\sim 110$ progenitor stars. In particular, the AGB models cover stellar masses from $1.3\ \rm M_\odot$ and $6.0\ \rm M_\odot$, and metallicities ranging from $Z=4.8\times10^{-5}$ to $Z=2.0\times10^{-2}$. Massive stars span progenitor masses from $13\ \rm M_\odot$ to $120\ \rm M_\odot$, with metallicities from $Z\simeq3.2\times10^{-5}$ to $Z\simeq1.3\times10^{-2}$ and three initial rotational velocities ($\mathrm{0,\ 150\ and\ 300\ km/s}$). With the current, state-of-the-art, prescriptions for s-process nucleosynthesis, some uncertainties still remain. In particular, results of chemical evolution models for different s-process elements seem to overestimate the solar abundance of those elements. This is often attributed to AGB stars (\citealp{Magrini2021, Vescovi2021}), as isotopic ratio measurements in presolar SiC grains suggest that the neutron densities in the FRUITY models may be overestimated (\citealp{Liu2018}). In chemical evolution models, this issue is typically addressed by applying a global reduction factor to the FRUITY yields for all neutron-capture elements (see e.g. \citealp{Rizzuti2019}). The overproduction is, however, not constant across mass, metallicity and different elements (see discussion in \citealp{Molero2025}). For this reason, in this work we adopt a more flexible, agnostic approach: the total s-process contribution (from both AGB stars and rotating massive stars) is selectively reduced for specific elements, when necessary, in order to match the s-process fractions of the solar system abundances reported in \citet{Prantzos2020}. A comparison between the s-process contribution to solar abundances predicted by our model and the s-process fractions estimated by \citet{Prantzos2020} is shown in Figure~\ref{fig: solar fraction}, together with the r- and p-process fractions and the total solar abundances from \citet{Lodders2009}. It is clear that reproducing the s-process fractions at solar metallicity does not necessarily imply that the full evolution of the s-process contributions is accurate across all metallicities. Nevertheless, the solar comparison consistency check, together with the advantage of using wide grids of stellar yields, provides some confidence in the modelled s-process evolution.

\subsection{R-process nucleosynthesis}

Classically, in both chemical evolution models and more complex cosmological magnetohydrodynamical simulations, the r-process is typically associated with one or more specific astrophysical sites, such as NSMs, MR-SNe or collapsars, each characterized by distinct delay times, event rates, nucleosynthetic yields and progenitor mass ranges (e.g., \citealp{Wehmeyer2015, vandevoort2020, vanderswaelmen2023}). These prescriptions are essential for testing theoretical models of r-process production against observational data, but they, by construction, impose assumptions about the nature of the contributing events. In this work, rather than committing to one or more specific astrophysical sites, we treat the r-process enrichment in a parametric way, so to explore a broader parameter space. The primary goal of this approach is not necessarily to identify a unique set of parameters that best fits the observational data, but rather to investigate whether there exists a range of conditions under which the enrichment history can simultaneously reproduce the observed trends of multiple neutron-capture elements.

Following the formalism of \citet{Matteucci2014}, the r-process contribution to the enrichment of a generic chemical element $i$ can be parametrized in chemical evolution models as:
\begin{equation}
R_{i,\rm r}(t) = Y_{i,\rm r} \times \alpha \int_{M_{\rm l}}^{M_{\rm u}} \psi(t - \tau_m - \tau) \phi(m) dm,
\label{eq: r-process rate}
\end{equation}
where $R_{i,\rm r}(t)$ is the rate at which r-process material in the form of the element $i$ is returned to the ISM, $Y_{i,\rm r}$ is the yield of the element $i$ per r-process event (in solar masses), and $\alpha$ is the fraction of stars in the mass range $[M_{\rm l}, M_{\rm u}]$ that lead to r-process production. The function $\psi(t)$ represents the star formation rate, $\phi(m)$ is the IMF, $\tau_m$ is the lifetime of a star of mass $m$, and $\tau$ is an additional delay time between the stellar death and the onset of r-process enrichment. This expression was originally developed to model the r-process contribution from NSMs, under the assumption that neutron star binary systems merge after a fixed timescale (it is worth noting that the update treatments of NSMs no longer assume a fixed delay time but instead adopt a delay-time distribution to account for the possible spread in the gravitational delay; see \citealp{Simonetti2019, Cote2019, Greggio2021}). In such a case, the term $\tau$ corresponds to the coalescence time of the binary system, and the mass range $[M_{\rm l}, M_{\rm u}]$ represents the birth masses of stars that can leave a neutron star as a remnant (above $\sim8\ \rm M_\odot$, consistent with the lower limit for CC-SNe progenitors adopted in the model). Similarly, if the extra delay time $\tau$ is set to zero, Equation \ref{eq: r-process rate} can represent r-process pollution by quick sources, generally attributed to a subclass of CC-SNe, as MR-SNe, which contribute to the early r-process enrichment without delay on top of that of the stellar lifetime $\tau_m$. Also in that case, $M_{\rm l}$ and $M_{\rm u}$ are greater than $\sim8\ \rm M_\odot$.

The relevant set of parameters explored in this work is $(Y_{\rm Eu, r},\ \alpha,\ [M_{\rm l}, M_{\rm u}],\ \tau)$. In particular, we vary the r-process yield of Eu, $Y_{\rm Eu,r}$, and derive the yields for the other neutron-capture elements of interest by assuming a fixed scaling relation between them and those of Eu. The adopted scaling factors are obtained from the solar system r-process contributions as determined by \citet{Simmerer2004}.
We run models spanning 20 values each of $(Y_{\rm Eu, r},\ \alpha,\ \tau)$, resulting in $8\ 000$ combinations. These are computed for 19 different choices of the progenitor mass range $[M_{\rm l},\ M_{\rm u}]$, for a total of $\sim 1.5\times 10^5$ model configurations. In particular:
\begin{itemize}
    \item $\alpha$ is linearly sampled in the range $[0,\ 1]$. Being the fraction of events that can successfully produce r-process material, by definition $\alpha=0$ corresponds to no r-process contribution and $\alpha=1$ corresponds to all events in the selected progenitor mass range contribute to the r-process enrichment.
    \item $Y_{\rm Eu,r}$ is logarithmically sampled in the range $[10^{-7},\ 10^{-5}]\ \rm M_\odot$. This range spans typical values for Eu production in NSMs as well as in MR-SNe (e.g., \citealp{Winteler2012, just2015, Nishimura2015, Nishimura2017, Reichert2021, Ricigliano2024}). The logarithmic sampling is adopted to ensure that the parameter space is explored uniformly in orders of magnitude.
    \item $\tau$ is sampled using a quasi-logarithmic distribution in the range $[0,\ 14]\ \rm Gyr$, with denser sampling at shorter timescales.
    \item $[M_{\rm l},\ M_{\rm u}]$ is sampled across the different intervals, starting from $[8,\ 10)\ \mathrm{M_\odot}$ and increasing in steps of $5\ \mathrm{M_\odot}$, up to $[95,\ 100]\ \mathrm{M_\odot}$.
\end{itemize}

The quality of each parameter combination is evaluated based on its ability to reproduce the observed abundance ratios trends of the neutron-capture elements of interest, using $\chi^2$ statistics as described in the following section.

\section{Model optimization}
\label{sec: Eu optimization}

Eu is one of the most reliable tracers of the r-process, with $\sim95\%$ of its solar abundance originating from rapid neutron captures (\citealp{Prantzos2020}). Its near-pure r-process origin, together with the quality of its observations (e.g., \citealp{Reddy2006, Hansen2012, Roederer2014, Lucertini2025}), makes it an ideal reference element for constraining the properties of the r-process enrichment in chemical evolution models. As described in section \ref{sec: s-process nucleosynthesis}, in this work we used prescribed s-process contribution for all neutron-capture elements and vary only the parameters controlling the r-process production: the event rate fraction $\alpha$, the yield per event $Y_{\rm Eu,r}$, the additional delay time $\tau$, and the progenitor mass range $[M_{\rm l}, M_{\rm u}]$.

\begin{figure*}
\begin{center}
    \subfloat{\includegraphics[width=0.5\textwidth]{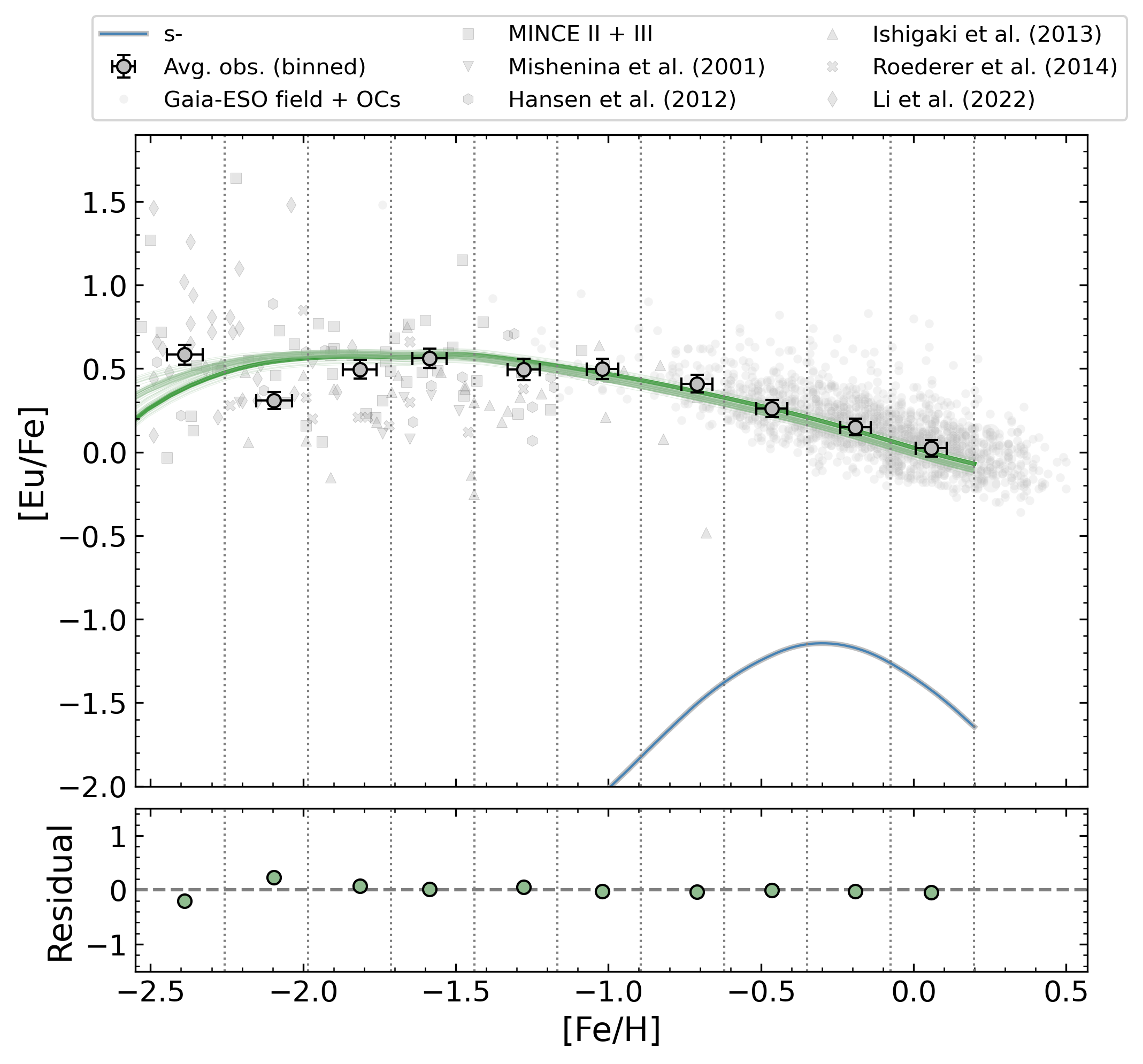}}
    \hfill
    \subfloat{\includegraphics[width=0.5\textwidth]{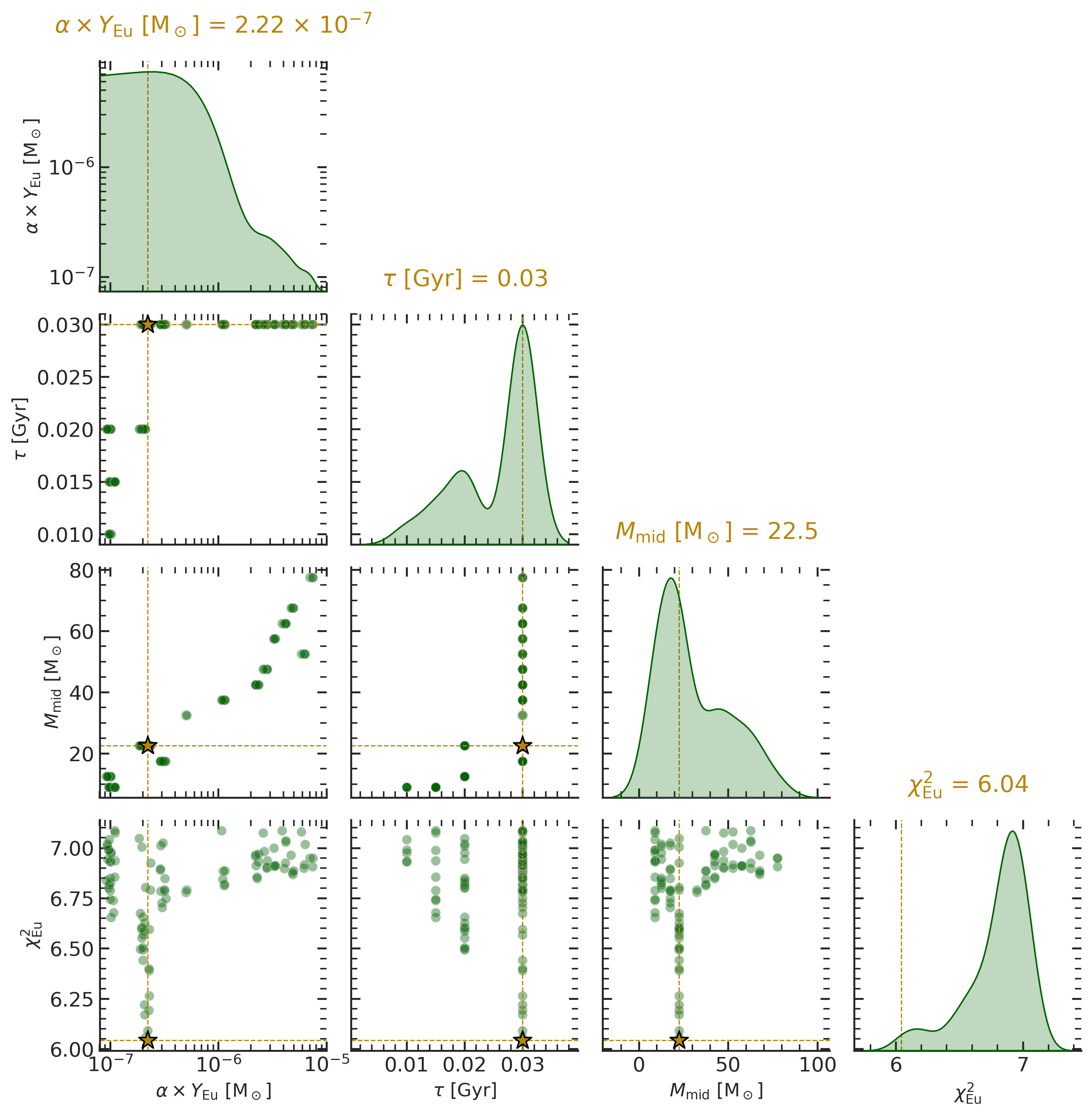}}
    \caption{Left panel: observed and predicted [Eu/Fe] vs. [Fe/H] trends. Grey circles with error bars show the average observational values in metallicity bins. The solid green curve shows the predicted [Eu/Fe] vs. [Fe/H] trend for the best-fitting model, while the surrounding lighter green lines represents the range spanned by models within the top 100, providing an indication of the model uncertainty. The lower, blue line represents the result of the model in case there is only the s-process contribution. The bottom panel shows the residuals between the binned observations and the model. Right panel: corner plot showing the marginal distributions (diagonal panels) and pairwise correlations (off-diagonal panels) of the parameters $(Y_{\rm Eu,\rm r},\ \alpha,\ [M_{\rm l}, M_{\rm u}],\ \tau)$ for the 100 models with the lowest $\chi^2_{\rm Eu}$. Dashed yellow lines indicate the values of the best-fitting model, with the corresponding numerical values reported above each marginal distribution, and the yellow star marks its position in each correlation plot.}
    \label{fig: best models EuFe vs FeH}
\end{center}
\end{figure*}

To quantify the agreement between model predictions and observations, we compute a binned $\chi^2$ for the [Eu/Fe] vs. [Fe/H] relation (see left panel in Figure \ref{fig: best models EuFe vs FeH} for reference). The metallicity range is divided into $N_{\mathrm{bins}}$ intervals of equal width and, for each bin $j$ containing both observational and model points, the weighted mean metallicity and abundance ratio of the observations $k$ and their uncertainties are computed as:
\begin{equation}
\langle X_i \rangle^{\rm obs}_j =
\frac{\sum_k w_{k,i}\,X_{k,i}}{\sum_k w_{k,i}}, 
\qquad 
w_{k,i} = \frac{1}{\sigma^2_{k,i}},
\qquad
\sigma^{\rm obs}_{j,i} = \sqrt{\frac{1}{\sum_k w_{k,i}}},
\end{equation}
where $i \in \{\mathrm{[Fe/H]}, \, \mathrm{[Eu/Fe]}\}$ and $\sigma_{k,i}$ are the observational uncertainties. The $\chi^2$ contribution in each bin $j$ is then:
\begin{equation}
\chi^2_j = \sum_i 
\left(
\frac{\langle X_i \rangle^{\rm obs}_j - 
      \langle X_i \rangle^{\rm mod}_j}{\sigma^{\rm obs}_{j,i}}
\right)^2 ,
\end{equation}
and the total $\chi^2$ is obtained as:
\begin{equation}
    \chi^2 = \frac{1}{N_{\rm bins}-N_{\rm free}}\sum_j\chi^2_j,
\end{equation}
where $N_{\rm free}$ is the number of free parameters varied in the model. The denominator thus represents the number of degrees of freedom. This $\chi^2$ formulation is adapted for comparing predictions from our homogeneous chemical evolution model with an observational dataset that may exhibit intrinsic scatter. By averaging the observations in metallicity bins, this statistic captures the underlying observed trend rather than being dominated by individual outliers, and accounts for the observational uncertainties in each bin. This prevents the model from being over-penalised for not reproducing stochastic abundance variations, such as those expected towards low metallicities, that lie outside the scope of homogeneous modelling. As discussed in section \ref{sec: observational data}, we exclude the very low-metallicity regime ($\mathrm{[Fe/H]} \leq -2.5\ \mathrm{dex}$) from our analysis to avoid the intrinsic stochasticity of the early chemical evolution. Nonetheless, some scatter remains in the data, even if to a much smaller degree and mainly confined to the lower end of the [Fe/H] range.

\subsection{Parameter sensitivity}
\label{sec: parameter sensitivity}

\begin{figure}
    \includegraphics[width=1\columnwidth]{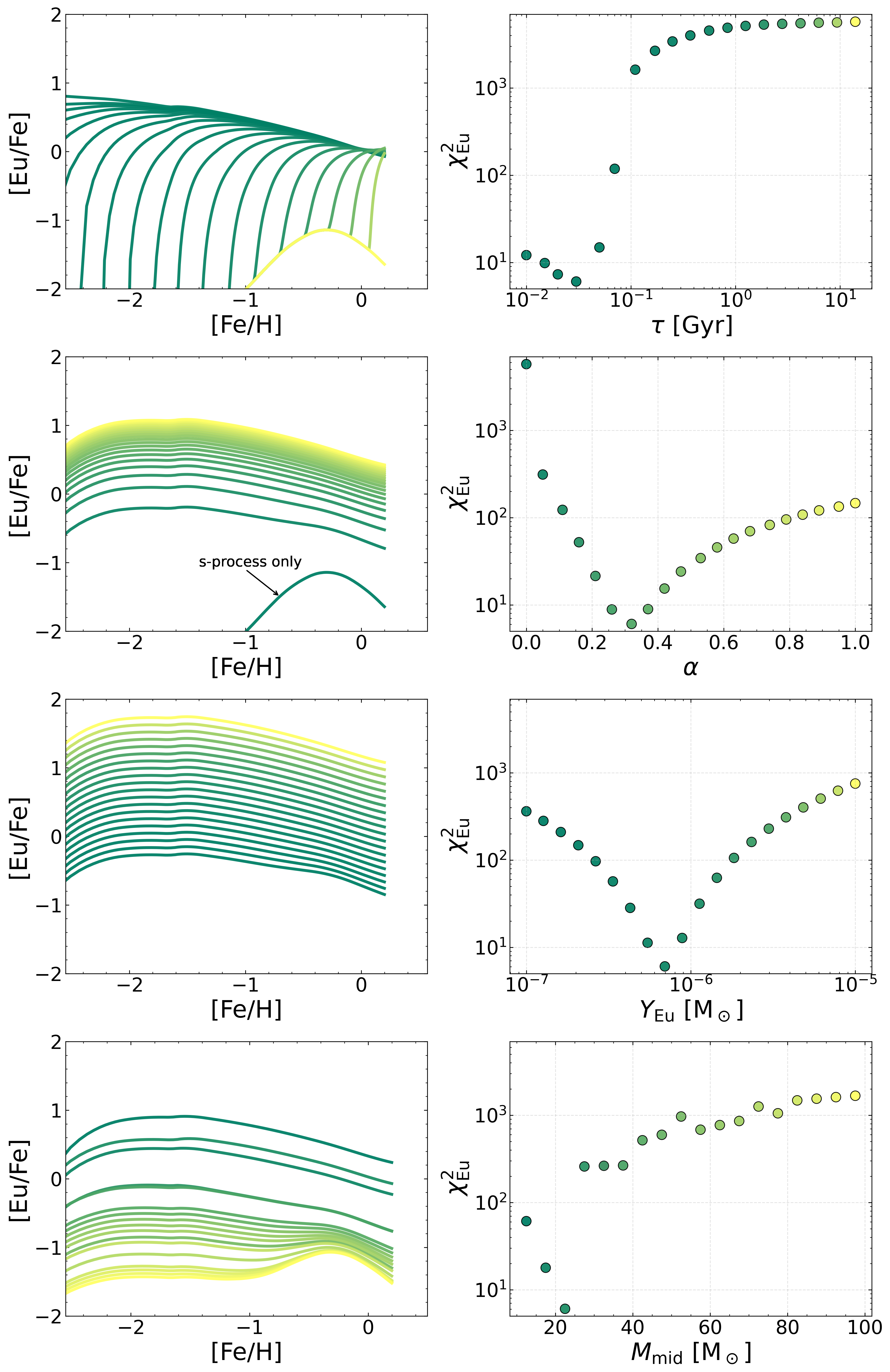}
     \caption{[Eu/Fe] vs. [Fe/H] predicted model curves (left column) for fixed combinations of three out of the four input parameters $(Y_{\rm Eu,\rm r},\ \alpha,\ [M_{\rm l}, M_{\rm u}],\ \tau)$, shown alongside the corresponding $\chi^2_{\rm Eu}$ values as a function of the fourth parameter (right column). In the left column, curve colours indicate the associated $\chi^2_{\rm Eu}$ values.}%
 \label{fig: test}%
\end{figure}

Before discussing the best-fit models for the [Eu/Fe] vs. [Fe/H] relation, it is interesting to first show how variations in our chosen input parameters influence the trend. Figure \ref{fig: test} presents this by showing, for fixed combinations of three out of the four input parameters $(Y_{\rm Eu,\rm r},\ \alpha,\ [M_{\rm l}, M_{\rm u}],\ \tau)$, the corresponding $\chi^2_{\rm Eu}$ values as a function of the fourth parameter, together with the resulting [Eu/Fe] vs. [Fe/H] curves. The chosen values are $\mathrm{\tau = 0.03\ \rm Gyr}$, $\alpha=0.32$, $Y_{\rm Eu, r}=6.95\times10^{-7}\ \rm M_{\odot}$ and $M_{\rm mid}=22.5\ \rm M_{\odot}$.

The effect of a longer time delay, $\tau$, is to shift the injection of Eu mass to later times, resulting in [Eu/Fe] curves that appear at progressively higher [Fe/H] values as $\tau$ increases. This effect in chemical evolution models is widely discussed in the context of whether NSMs can be the sole source of Eu (e.g., \citealp{Argast2004, Matteucci2014, Cescutti2015, Schonrich2019}). The late injection of r-process material, consistent with longer merger delays driven by gravitational-wave emission, makes it difficult to reproduce the observed plateau in the [Eu/Fe] vs. [Fe/H] relation towards lower metallicities, highlighting the need for at least one additional r-process production site. Generally, difficulties emerge not only at low metallicities, but also at high [Fe/H], where the late injection of Eu counterbalances the delayed Fe enrichment from Type Ia SNe, producing a rise in the predicted curves which is not usually observed. In this work, we also explore long time delays ($\mathrm{>10\ Gyr}$), which produce curves where this effect is visible (see upper left panel of Figure \ref{fig: test}), with the longest time delay consequently leading to higher $\chi^2_{\rm Eu}$ values. This preference for shorter delay times, and thus faster r-process sources, is evident here as well, with the four smallest $\tau$ values yielding the lowest $\chi^2_{\rm Eu}$.

Regarding the fraction of events that end up producing r-process material, $\alpha$, its effect is relatively straightforward: increasing $\alpha$ corresponds to higher [Eu/Fe] vs. [Fe/H] curves. $\alpha = 0$, which implies no r-process enrichment (see equation \ref{eq: r-process rate}), yields the highest $\chi^2_{\rm Eu}$ value. This case corresponds to the low [Eu/Fe] vs. [Fe/H] curve starting at $\mathrm{[Fe/H]\simeq-1\ dex}$, produced when only the s-process is active. 

A similar effect to that of $\alpha$ is obtained by varying the Eu yield, $Y_{\rm Eu}$. These two parameters are indeed, to some extent, degenerate, and we will therefore often refer to their product, rather than to their individual values (quantity that we indicate as the effective Eu injection rate).  

Finally, increasing the progenitor mass range $[M_{\rm l}, M_{\rm u}]$ produces a downward shift of the [Eu/Fe] vs. [Fe/H] curves, as an effect of the IMF weighting that favours the formation of lower-mass stars. Higher mass progenitors tend to be disfavoured, with the $\chi^2_{\rm Eu}$ values generally increasing as the mass range is shifted toward higher masses, although the minimum $\chi^2_{\rm Eu}$ does not corresponds to the lowest mass range. The trend is not strictly monotonic, as some mass ranges correspond to higher $\chi^2_{\rm Eu}$ values than neighbouring ranges. 

We note that, once all combinations of the input parameters are explored, the resulting models sample the full [Eu/Fe] vs. [Fe/H] parameter space, allowing a complete analysis with the results not being biased by specific parameter choices.

\subsection{Best-Fit models}
\label{sec: best-fit models}

From the $\sim 1.5 \times 10^5$ models computed, we select the 100 with the lowest $\chi^2_{\rm Eu}$. The corresponding parameter configurations and predicted [Eu/Fe] vs. [Fe/H] trends are shown in Figure \ref{fig: best models EuFe vs FeH}. Note that, in the pairwise correlation panels (off-diagonal panels in the corner plot), we show the effective Eu injection rate $\alpha \times Y_{\rm Eu}$, because of the previously noted parameter degeneracy between $\alpha$ and $Y_{\rm Eu}$. 

The best models cluster at short delay times, with $\tau$ not exceeding $30\ \mathrm{Myr}$. Even when added to the lifetime of the least massive progenitor considered in the model ($8\ \rm M_\odot$, with lifetime $\tau_{\rm m} \simeq 55\ \mathrm{Myr}$ at solar metallicity), the resulting enrichment still occurs on sufficiently short timescales. This ensures an almost immediate injection of r-process material into the ISM, consistent with the requirement of rapid enrichment to match the low-metallicity observations. Moreover, since in our parametrization each model includes only a single r-process source, this is consistent with the idea that when a rapid source contributes to Eu enrichment alongside a delayed one, the rapid source dominates the overall enrichment (\citealp{vanderswaelmen2023, Molero2023}).  As already mentioned in the previous section, the four shortest $\tau$ values are therefore the preferred ones, providing similar results, with differences only at very low [Fe/H] values (as shown in the previous section, see upper panels of Figure \ref{fig: test}). Finally, although the distribution of the best $\tau$ values appears bimodal in the pairwise correlation panel of Figure \ref{fig: best models EuFe vs FeH}, we note that the interval between $20$ and $30\ \mathrm{Myr}$ is not sampled.

The progenitor mass range does not exceed $[75,\ 80)\ \rm M_\odot$, with this upper limit corresponding to the highest values of $\alpha \times Y_{\rm Eu}$. This follows the effect of the IMF, which favours the formation of lower-mass progenitors suppressing the contributions from the most massive stars. The best models preferentially select progenitors in the $[20,\ 25)\ \rm M_\odot$ range. However, as shown by the $M_{\rm mid}$ vs. $\alpha \times Y_{\rm Eu}$ correlation, a degeneracy exists between these two parameters: increasing the progenitor mass reduces the number of stars formed, which must be compensated by a higher Eu yield. As a result, different combinations of mass and yield can reproduce similar [Eu/Fe] vs. [Fe/H] trends, and the preferred $[20,\ 25)\ \rm M_\odot$ range should not be regarded as a too strict constraint, but rather as an indication of a preference toward the lower-mass regime.

The result of best-fitting model, with $\chi^2_{\rm Eu} \simeq 6$, is represented by the solid green line (left panel in Figure \ref{fig: best models EuFe vs FeH}). This $\chi^2_{\rm Eu}$ value indicates a good fit, though some margin for improvement remains. The agreement is in fact stronger towards higher metallicities ($\mathrm{[Fe/H] \gtrsim -1.7\ dex}$; see residual panel), where the observational data exhibit indeed less scatter, reflecting a more homogeneous chemical evolution. The relatively high value of the $\chi^2_{\rm Eu}$ is indeed mainly due to its values in the two lowest metallicity bins, equal to $\chi^2_{\rm Eu} \simeq 11$ for the range $-2.53 \leq \mathrm{[Fe/H]} \leq -2.26$ and to $\chi^2_{\rm Eu} \simeq 19$ for the range $-2.25 < \mathrm{[Fe/H]} \leq -1.98\ \rm dex$, while being $< 1$ for higher [Fe/H]. Nevertheless, the best 100 models span a narrow range of $\chi^2$ values, with a maximum $\chi^2_{\rm Eu}$ only slightly higher than that of the best model (i.e. $\chi^2_{\rm Eu}\simeq7$), indicating that the region around the optimal solution is relatively well constrained.

\begin{figure*}
    \includegraphics[width=1\textwidth]{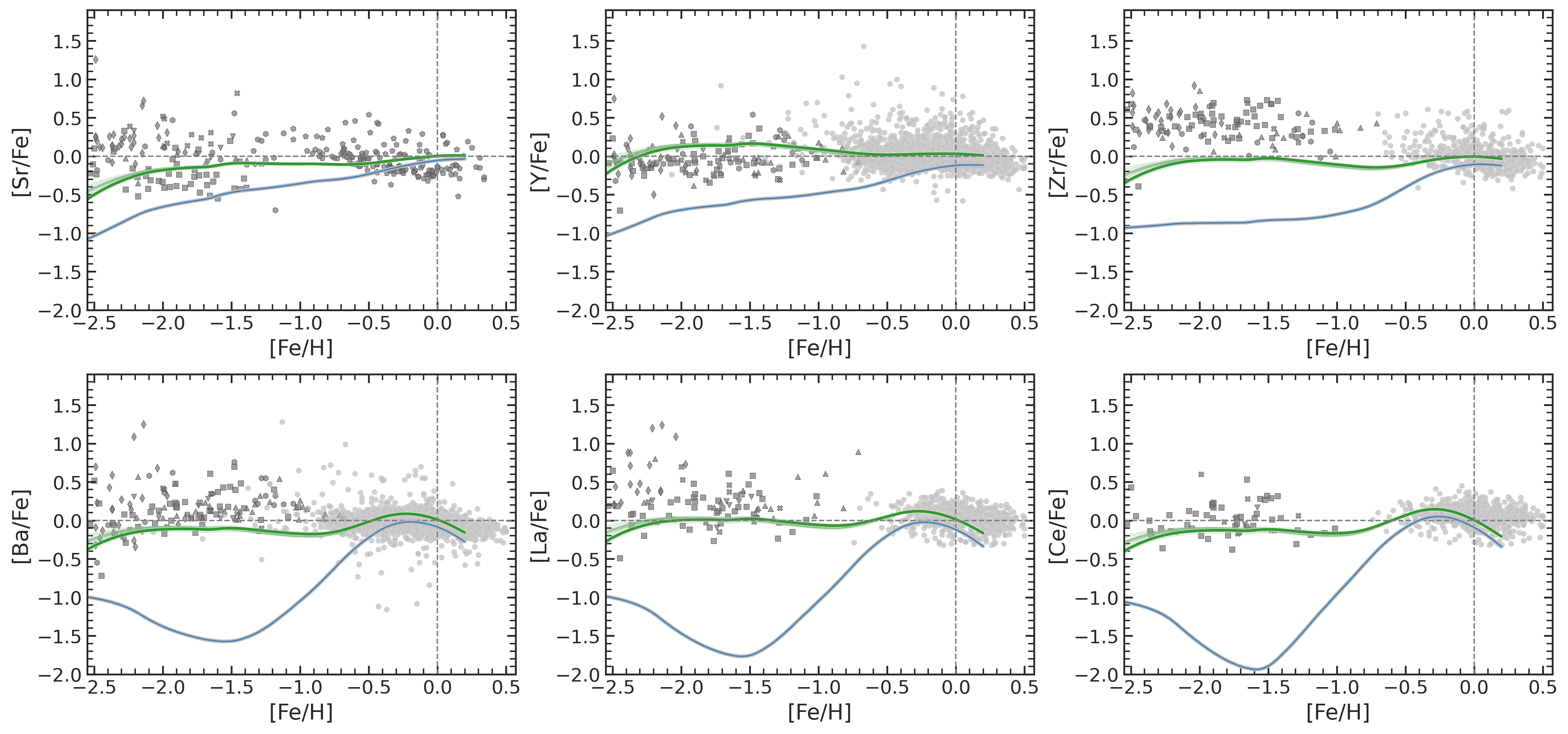}
     \caption{Observed and predicted [X/Fe] vs. [Fe/H] trends for Sr, Y, Zr and Ba, La, Ce. Solid green curves shows the predicted trend for the best-fitting [Eu/Fe] vs. [Fe/H] trend, while lighter green lines represents that range spanned by models within the top 100. The lower, light blue lines represent the results of the model in case there is only the s-process contribution. Symbols for the observational data are the same as in Figure \ref{fig: best models EuFe vs FeH}.}%
 \label{fig: ncFe_1st_2nd_FeH}%
\end{figure*}

In the chemical evolution model, the r-process yields for other neutron-capture elements of interest are derived from the Eu yield by scaling them according to the solar r-process contribution. The underlying assumption is the existence of a metallicity-independent scaling relation between the Eu r-process yield and that of all other neutron-capture elements. This implicitly extends the concept of r-process universality, typically observed between the lanthanides and the third-peak neutron-capture elements, to lighter elements as well (see \citealp{Roederer2022} for a discussion). While this can be a strong assumption, it is commonly adopted in chemical evolution models (e.g. \citealp{Cescutti2014, Rizzuti2019, Molero2021, Palla2025}). Nevertheless, if such an assumption holds, one would expect the best-fitting Eu models to also reproduce the observed [X/Fe] vs. [Fe/H] trends of the lighter neutron-capture elements. Moreover, provided that the s-process contribution is modelled accurately, this agreement should hold not only at solar metallicity, but across the entire metallicity range. The results of the best-fitting Eu models for the lighter neutron-capture elements Sr, Y, Zr and for Ba, La, Ce are shown in Figure \ref{fig: ncFe_1st_2nd_FeH}. For all elements, the solar [X/Fe] is well reproduced. However, while at higher metallicities the best-fitting Eu models show generally good agreement with the observations, at lower [Fe/H] the observed trends are reproduced only in some cases. Among the lightest elements, the models reproduce the [Y/Fe] vs. [Fe/H], but systematically underestimate the [Sr/Fe] and [Zr/Fe], with the largest discrepancy found for Zr. The agreement is generally better for Ba, La, and Ce, although the models still tend to underestimate the observed [Ba/Fe] and [La/Fe] trends. This behaviour may indicate i) a non-universality of the r-process for the lighter neutron-capture elements, as the better agreement obtained by the models for Ba, La and Ce suggests that the solar-scaled relation is a more robust assumption for heavier elements, consistent with them being less influenced by additional processes (\citealp{Cowan2021}); ii) a possible metallicity-dependent scaling between Eu and other r-process elements; and, although not addressed here, iii) uncertainties in the adopted s-process yields. 

In order to investigate whether a simultaneous fit to different neutron-capture elements can be still achieved, a multi-objective optimisation based on a Pareto-front algorithm is presented in section \ref{sec: Multi-objective optimization}. Such an approach can allow us to find possible trade-offs in reproducing the observed trends for different elements and to evaluate whether it is still possible to reproduce the full set of observed abundance ratios patterns across all elements without additional nucleosynthetic channels. In the next section, we give a general introduction to the concept of multi-objective optimization and the adopted Pareto front algorithm, but for a more detailed description we refer to \cite{Ngatchou2005} (see also \citealp{Ehrgott2025} for a recent review).

\section{Multi-objective optimization}
\label{sec: Multi-objective optimization}

In the previous section, we assess the agreement between model predictions and observational data using a single diagnostic, namely the reduced $\chi^2$ statistic for [Eu/Fe] vs. [Fe/H]. While this provides a straightforward measure of the quality of the model in reproducing the Eu, it implicitly assumes that one observable is enough to characterize the full parameter space, neglecting potential trade-offs between different constraints. 

If the interest is in a single solution that simultaneously fit different neutron-capture elements patterns, multiple objectives (fit of multiple elements) can be converted into a single objective function by aggregating all of them in a weighted function. However, this approach requires an a priori knowledge on the weights that reflect the relative importance of each observable. Moreover, and more importantly for the scope of this work, such a function converts the multi-dimensional problem into a single objective one, making it difficult to evaluate whether a compromise between different, and often conflicting, objectives exists. 

Since we also desire to know all possible optimization solutions of all objectives at the same time (a trade-off analysis), a single optimization process is not suited for our case. Instead, we frame the problem within the context of multi-objective optimization, where the aim is not to find a single global minimum of a scalar cost function, but to find the set of solutions that achieve the best compromise among all the different objectives. 

A multi-objective optimization problem can be written in a general form as:
\begin{equation}
\begin{aligned}
    & \text{minimize}   && \boldsymbol{F}(\mathbf{x}) = [f_1(\mathbf{x}), f_2(\mathbf{x}), \ldots, f_m(\mathbf{x})], \\
    & \text{subject to} && \mathbf{x} \in \Omega ,
\end{aligned}
\end{equation}
where $\mathbf{x}$ is a vector of decision variables within the parameter space $\Omega$, $f_i(\mathbf{x})$ ($i=1,\dots,m$) are the objective functions to be minimized, and $m>1$ indicates that the problem is multi-objective. In our case, the decision variables $\mathbf{x}$ correspond to the chemical evolution model parameters $(Y_{\rm Eu,\rm r},\ \alpha,\ [M_{\rm l}, M_{\rm u}],\ \tau)$. For each $\mathbf{x}$, we compute a vector of reduced $\chi^2$ values,
\begin{equation}
\boldsymbol{F}(\mathbf{x}) = \big[\chi^2_{1}(\mathbf{x}), \chi^2_{2}(\mathbf{x}), \ldots, \chi^2_{m}(\mathbf{x})\big],
\end{equation}
where $m$ is the number of elemental abundance trends included in the analysis, so that each component of this vector quantifies the agreement between the model prediction and the corresponding observational trend of a given element $i$.

In general, if $m=1$, the problem reduces to the standard single-objective case, which typically admits a unique optimal solution. On the other hand, when $m>1$ and the different objectives are in conflict, no single solution simultaneously minimizes all of them. Instead, one generally obtains a set of optimal compromises, the so-called non dominated solutions.

\subsection{The Pareto front algorithm}

Solutions to the multi-objective problem can be classified as dominated or non-dominated. A solution $\mathbf{x}_A \in \Omega$ is said to dominate another solution $\mathbf{x}_B \in \Omega$ if, in the minimization context:
\begin{equation}
\begin{aligned}
& f_i(\mathbf{x}_A) \leq f_i(\mathbf{x}_B) \quad \forall i \in { 1,\dots,m }, \\
& \text{and} \quad \exists j \in { 1,\dots,m }:\ f_j(\mathbf{x}_A) < f_j(\mathbf{x}_B).
\end{aligned}
\end{equation}
That is, $\mathbf{x}_A$ dominates $\mathbf{x}_B$ if $\boldsymbol{F}(\mathbf{x}_A)$ is no worse than $\boldsymbol{F}(\mathbf{x}_B)$ in all objectives and strictly better in at least one. A solution that is not dominated by any other is called Pareto optimal and, by definition, cannot be improved in one objective without worsening at least one other objective. The set of all Pareto-optimal solutions is called the Pareto set, and their image in the objective space defines the Pareto front. In two-dimensional objective space, the Pareto front appears as the lower-left envelope of the distribution of models (or upper-right in maximization context), while in three dimensions it forms a surface. For higher dimensions, the concept generalizes directly, although it is less intuitive to visualize, as we will describe in the next sections.

The Pareto front thus might provide the boundary of feasible trade-offs, since every solution outside this boundary is strictly sub-optimal and every solution on the front represents a model that cannot be improved for one observable without worsening another one. Among these solutions, we will also show a single representative model on the Pareto front that minimized the Euclidean distance to the point $\mathbf{0}$ in objective space, $D$.

\section{Results}
\label{sec: results}

In this section, we present the results of the parameter-space exploration and multi-objective optimization analysis. For each chemical evolution model, we compute the reduced $\chi^2$ values quantifying the agreement with the observed [X/Fe] vs. [Fe/H] trends, and apply the Pareto front formalism to identify the set of models that represent optimal trade-offs among different elemental constraints. The results are organized by element groups, we first present the Eu–Zr pair, which forms the two-dimensional case study for our multi-objective approach. We then extend the analysis to other neutron-capture elements, representing both the three- and multi-dimensional cases, examining whether the best-fitting Eu models also reproduce the observed patterns of first- and second-peak s-process elements. We are interested both in the regions of the parameter space that provide a good simultaneous fit, and to those that might highlight the limitations of assuming simple solar yield-scaling relations.

\subsection{Eu \& Zr}

The Eu-Zr pair represents the simplest two-dimensional case of our analysis. Eu is used as the canonical r-process tracer element that we adopt for our first model calibration (see section \ref{sec: Eu optimization}). Zr, on the other hand, represents a challenging case, since as shown in Figure \ref{fig: ncFe_1st_2nd_FeH}, its [Zr/Fe] trend does not follow the observations when its abundance is scaled assuming a solar pattern from the optimized Eu abundance. 

\begin{figure}
    \includegraphics[width=0.9\columnwidth]{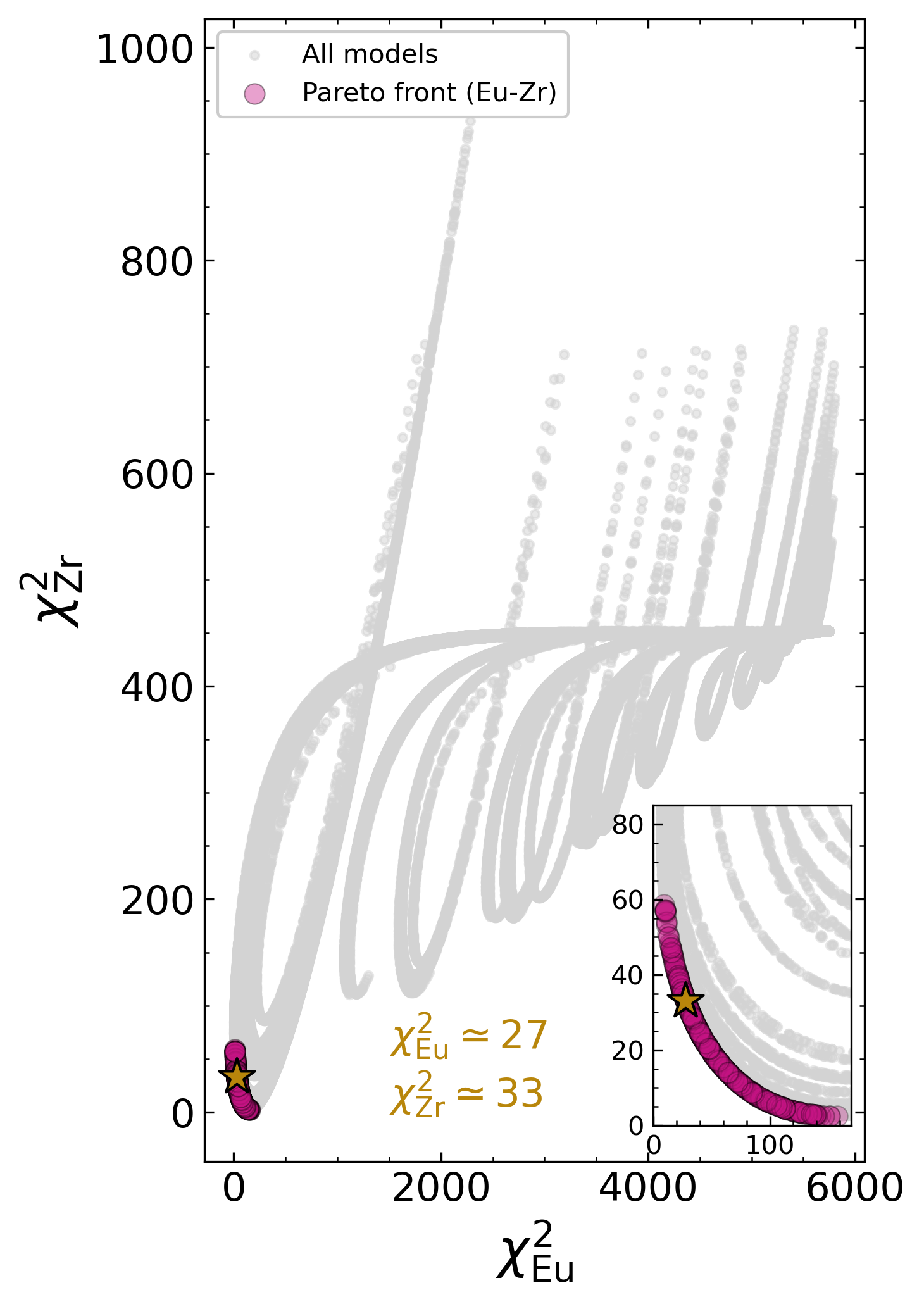}
     \caption{Reduced $\chi^2$ values for [Eu/Fe] vs. [Fe/H] and [Zr/Fe] vs. [Fe/H]. Grey points show all computed models, while magenta points highlight the Pareto-optimal solutions for the Eu–Zr pair, corresponding to the lower-left boundary of the objective space. The zoomed plot shows the low–$\chi^2$ region, where the optimal compromise between $\chi^2_{\mathrm{Eu}}$ and $\chi^2_{\mathrm{Zr}}$ is located (yellow star).}%
 \label{fig: chi2 Pareto front}%
\end{figure}

In this two-objective case, the quantities to be minimized are the reduced $\chi^2$ values of the [Eu/Fe] vs. [Fe/H] and [Zr/Fe] vs. [Fe/H] relations ($\chi^2_{\mathrm{Eu}}$ and $\chi^2_{\mathrm{Zr}}$, respectively). Figure \ref{fig: chi2 Pareto front} shows the full distribution of the $\chi^2$ values together with the corresponding Pareto-optimal models. All the $\chi^2$ results cluster along distinct J-shaped curves, a structure that reflects variations in the input parameters $(Y_{\rm Eu,\rm r},\ \alpha,\ [M_{\rm l}, M_{\rm u}],\ \tau)$. 
In particular, the distinct J-shaped families of solutions primarily reflect variations in the delay time $\tau$: overall each J  corresponds to a different $\tau$, and the general agreement with the data worsens systematically as $\tau$ increases, shifting the curves upward and to the right. Within a given J, the location along the diagonal branch is controlled by the effective Eu injection rate, $\alpha \times Y_{\rm Eu}$. Moving down the branch towards lower $\alpha \times Y_{\rm Eu}$ improves the simultaneous fit to Eu and Zr, but further reduction eventually worsens the agreement again. Finally, variations in the $M_{\rm mid}$ create secondary effects in the overall structure of the curve, which might however not be immediately apparent, since  $M_{\rm mid}$ is partially degenerate with both $\tau$ and with the Eu injection rate. Indeed, changing  the progenitor mass changes the stellar lifetime and, moreover, the number of contributing progenitors depends on the IMF weighting.

The Pareto front models identified for Eu and Zr provide a first attempt to test if the parameter configurations that successfully reproduce Eu can simultaneously fit the [Zr/Fe] vs. [Fe/H] trend. As described in the previous section, the Pareto front in two dimensions corresponds to a curve located in the lower-left region of the objective space, where simultaneous improvements in both objectives are no longer possible. This is indeed visible in our results (see magenta dots in Figure \ref{fig: chi2 Pareto front}).

The properties of these Pareto-optimal Eu–Zr models are shown in Figure \ref{fig: pairplot Pareto models Eu-Zr} with the parameters correspondent to the best Pareto model summarized in Table \ref{tab: model results}. As in the case of Eu, $\tau$ remains confined to very short values ($\tau \lesssim 30$ Myr), consistent with the requirement of rapid enrichment to match the low-metallicity data. The distribution of $M_{\rm mid}$ indicates a preference for progenitors in the lower part of the explored mass range, with no models with progenitors more massive than $80\ \mathrm{M_\odot}$ belonging to the Pareto ones. As discussed in section \ref{sec: best-fit models}, this is due to the IMF weighting: there are not enough stars with initial mass higher than that value to successfully reproduce the observed trends, even when assuming high r-process injection rates. As underlined in section \ref{sec: Eu optimization}, because of the strong degeneracy between $Y_{\rm Eu}$ and $\alpha$, here we report their product distribution, rather than their single values. It is interesting to note that the parameter values correspondent to the model closest to the point $(0,0)$ in the objective space are very similar to those of the Eu-only best-fit model (see Table \ref{tab: model results}), differing only by a moderately higher Eu injection rate (still within the same order of magnitude) and a slightly shorter delay time $\tau$. 

\begin{table}[]
    \centering
    \caption{Best Pareto-optimal model parameters for the different element groups. The Eu injection rate, $\alpha \times Y_{\rm Eu}$, the delay time, $\tau$, the medium value of the progenitor mass range, $M_{\rm mid}$, and the distance to the point \textbf{0} in objective space, $D$.}
    \begin{tabular}{c|c|c|c|c}
    \hline
    \hline
         Model & $\alpha \times Y_{\rm Eu}\ [\rm M_\odot]$ & $\tau\ [\rm Gyr]$ & $M_{\rm mid}\ [\rm M_\odot]$ & $D$\\
         \hline
         Eu          & $2.22 \times 10^{-7}$ & 0.03 & 22.5 & 6.0  \\
         Eu-Zr       & $2.94 \times 10^{-7}$ & 0.01 & 22.5 & 43.0 \\
         Sr-Y-Zr     & $3.44 \times 10^{-7}$ & 0.02 & 22.5 & 37.7 \\
         Ba-La-Ce    & $1.78 \times 10^{-7}$ & 0.01 & 9.0  & 9.9  \\
         Eu-Sr-Y-Zr  & $3.01 \times 10^{-7}$ & 0.02 & 22.5 & 47.4 \\
         Eu-Ba-La-Ce & $2.70 \times 10^{-7}$ & 0.02 & 22.5 & 23.3 \\
    \hline
    \end{tabular}
    \label{tab: model results}
\end{table}

\begin{figure*}
    \includegraphics[width=1\textwidth]{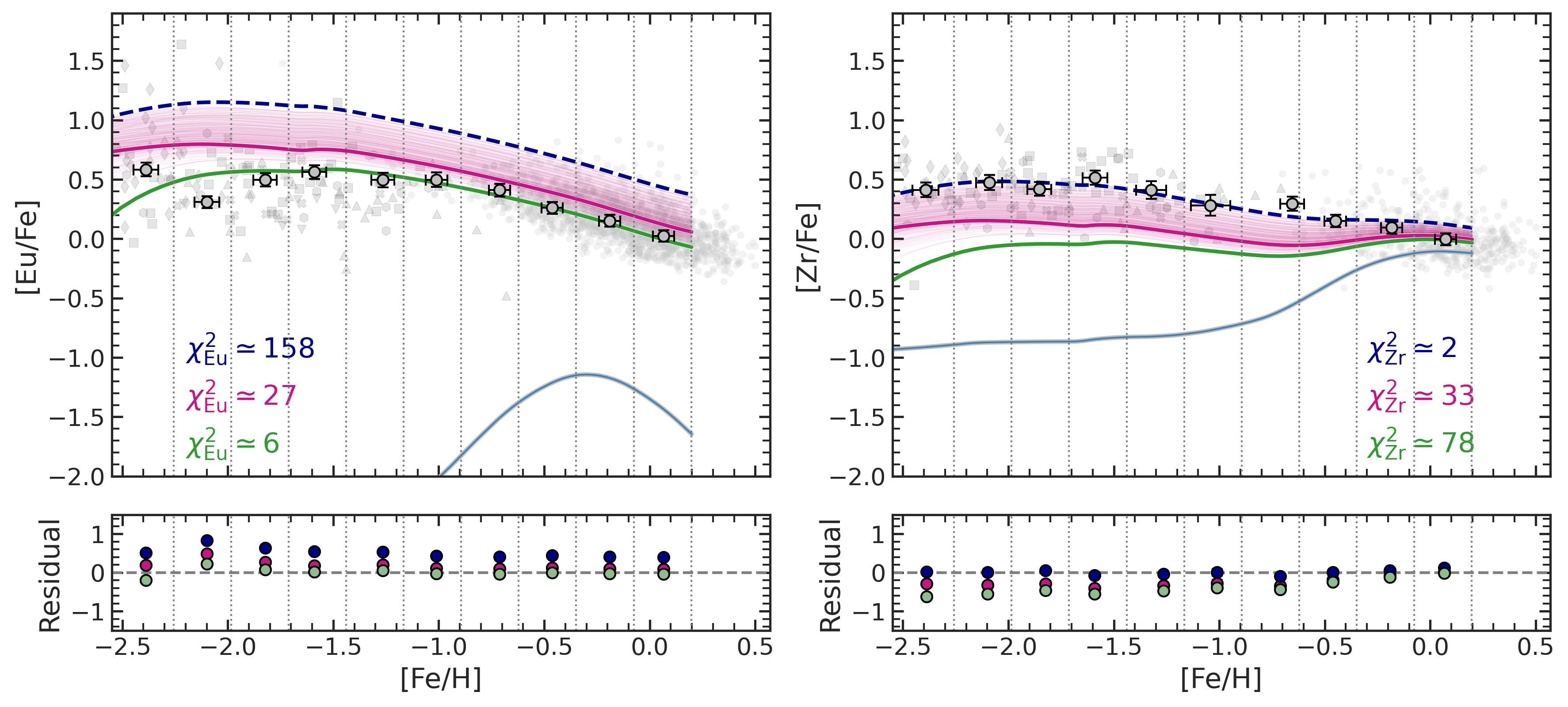}
     \caption{Comparison between observed [Eu/Fe] and [Zr/Fe] vs. [Fe/H] trends and the predictions of the Pareto optimal models (magenta lines). The thicker magenta lines show the predicted [Eu/Fe] and [Zr/Fe] vs. [Fe/H] trends for the Pareto optimal model with the smallest distance to the ideal point $(0,0)$ in objective space. The solid green curve shows the best-fit model to Eu and its corresponding prediction for Zr, while the dashed blue line represents the best-fit model to Zr and its prediction for Eu. The lower, light blue lines represent the results of the model in case there is only the s-process contribution. Grey circles with error bars show the average observational values in metallicity bins. Residuals relative to the binned observations are displayed in the lower panels. Symbols for the observational data are the same as in Figure \ref{fig: best models EuFe vs FeH}.}%
 \label{fig: Eu_Zr_FeH_pareto}%
\end{figure*}

The results of the Pareto optimal models for the [Eu/Fe] and [Zr/Fe] vs. [Fe/H] are represented in Figure \ref{fig: Eu_Zr_FeH_pareto} by the magenta lines. As expected from the coupling seen in the objective space, an increase in the effective r-process injection rate simultaneously raises both the Eu and Zr abundances. At first, this leads to an improved agreement with the data for both elements, but further increases in the injection rate eventually worsen the fit. The resulting Pareto solutions cluster into a band that, at least qualitatively, reproduces the overall behaviour of both [Eu/Fe] and [Zr/Fe]. However, the [Zr/Fe] is well matched by those Pareto solutions that overestimate the [Eu/Fe], whereas those models that best reproduce Eu systematically underpredict Zr. This is further highlighted by the individual best-fit models for Eu and Zr and their corresponding predictions for the other element (green and blue curves, respectively). The best Eu model, characterised by a low $\chi^2_{\rm Eu}$, yields a [Zr/Fe] trend that lies at the lower edge of the Pareto band and provides the worst agreement with the Zr data (high $\chi^2_{\rm Zr}$). In contrast, the best Zr model, which minimises $\chi^2_{\rm Zr}$, corresponds to a severe overestimation of [Eu/Fe], producing the largest $\chi^2_{\rm Eu}$ among the Pareto solutions. The Pareto optimal model closest to the point $(0,0)$ in the objective space (thicker magenta lines), provides intermediate values of $\chi^2_{\rm Eu}$ and $\chi^2_{\rm Zr}$. It reproduces the [Eu/Fe] trend more successfully than the [Zr/Fe] and, overall, shows a better agreement towards higher metallicities.

\subsection{Light and heavy neutron-capture elements}
\label{sec: Light and heavy neutron-capture elements}

A natural test of the r-process contribution would be that of comparing model predictions directly to elements belonging to the r-process peaks, as well as the lanthanides. In our model, these include species such as Se, Br and Kr for the first peak, Te, I and Xe for the second peak, Gd, Tb and Dy for the Lanthanides, and Os, Ir and Pt for the third r-process peak (see Figure \ref{fig: solar fraction} for reference). However, in practice, observational data for all the r-process elements (excluding Eu) are too few to permit a meaningful statistical comparison, remaining limited to small stellar samples. We therefore focus our statistical investigation on the more extensively observed neutron-capture elements: Sr, Y and Zr for the region between the first and second r-process peaks, and Ba, La and Ce for the region beyond the second peak. Although these elements are classified as s-process elements, their abundances towards low metallicity reflect the r-process enrichment, with the s-process becoming dominant only towards higher metallicities (see, e.g., blue lines in Figure \ref{fig: ncFe_1st_2nd_FeH}). Moreover, their broad observational coverage across the full [Fe/H] range makes them well suited for comparison with the chemical evolution models and, in particular, for our statistical investigation on the universality of the r-process as a function of both elements and metallicity. 

Nevertheless, the few observational data available for the r-process elements will still be adopted for a qualitative comparison with our model predictions. These elements are therefore not included in the statistical optimisation, but provide a secondary check of the model results.

\subsubsection{Sr, Y, Zr}
\label{sec: Sr, Y, Zr}

The case of Sr, Y and Zr provides the extension of our two-dimensional Eu-Zr analysis into a three-dimensional case. Within our modelling approach, their r-process yields are scaled from Eu according to the solar r-process residuals. This procedure, as seen in section \ref{sec: best-fit models} (see Figure \ref{fig: ncFe_1st_2nd_FeH}), leads to a good prediction of the [Y/Fe] vs. [Fe/H] pattern, but it systematically underestimates the observed [Sr/Fe] and [Zr/Fe] ratios, with Zr being the most critical case. By studying Sr, Y and Zr simultaneously, we construct a three-dimensional objective space that allow us to asses whether consistent parameter configurations can be found that reproduce all three elements together. Moreover, if such an investigation is done independently of Eu, we can test whether the first-peak elements alone favour a distinct region of parameter space. 

\begin{figure*}
\begin{center}
    \subfloat{\includegraphics[width=0.5\textwidth]{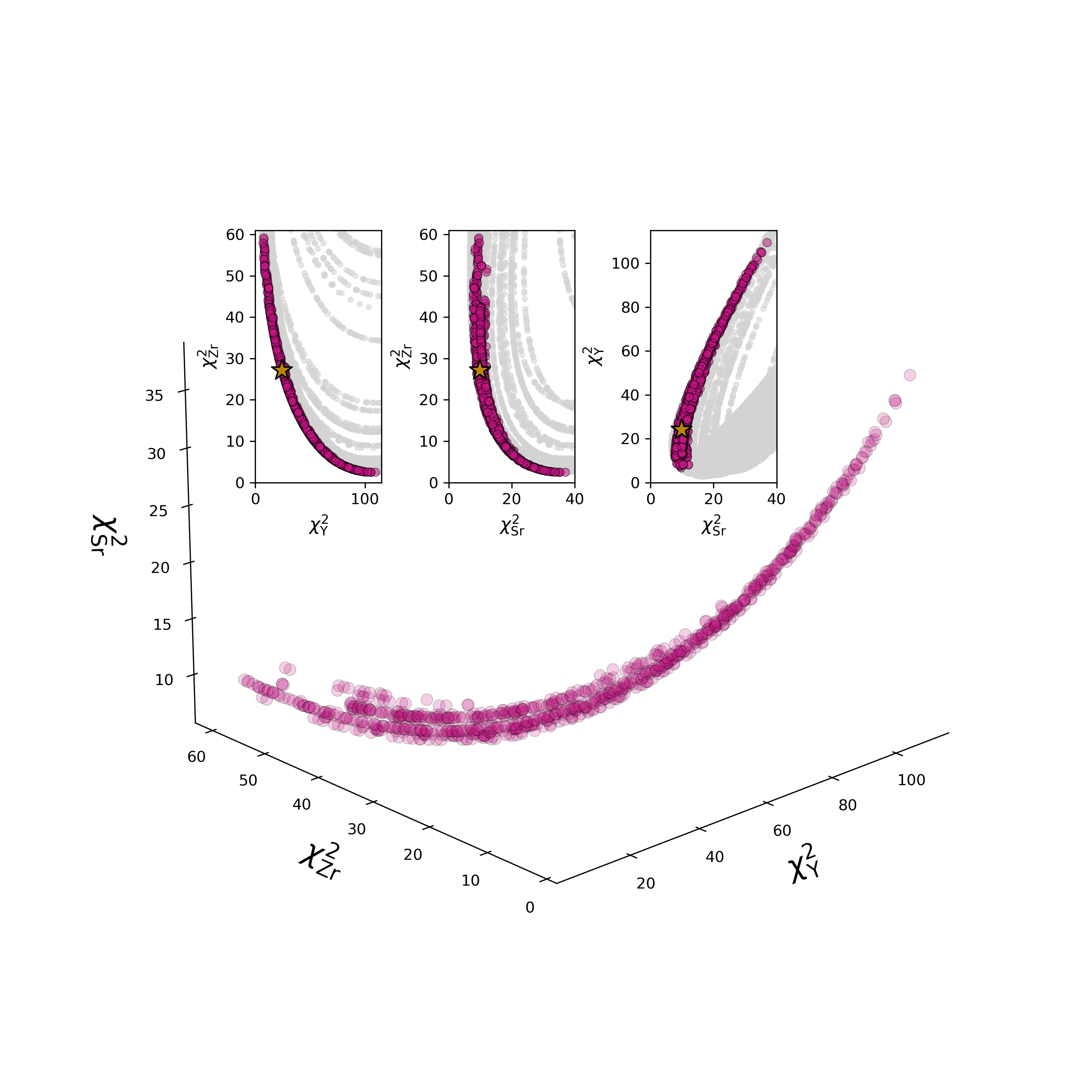}}
    \hfill
    \subfloat{\includegraphics[width=0.5\textwidth]{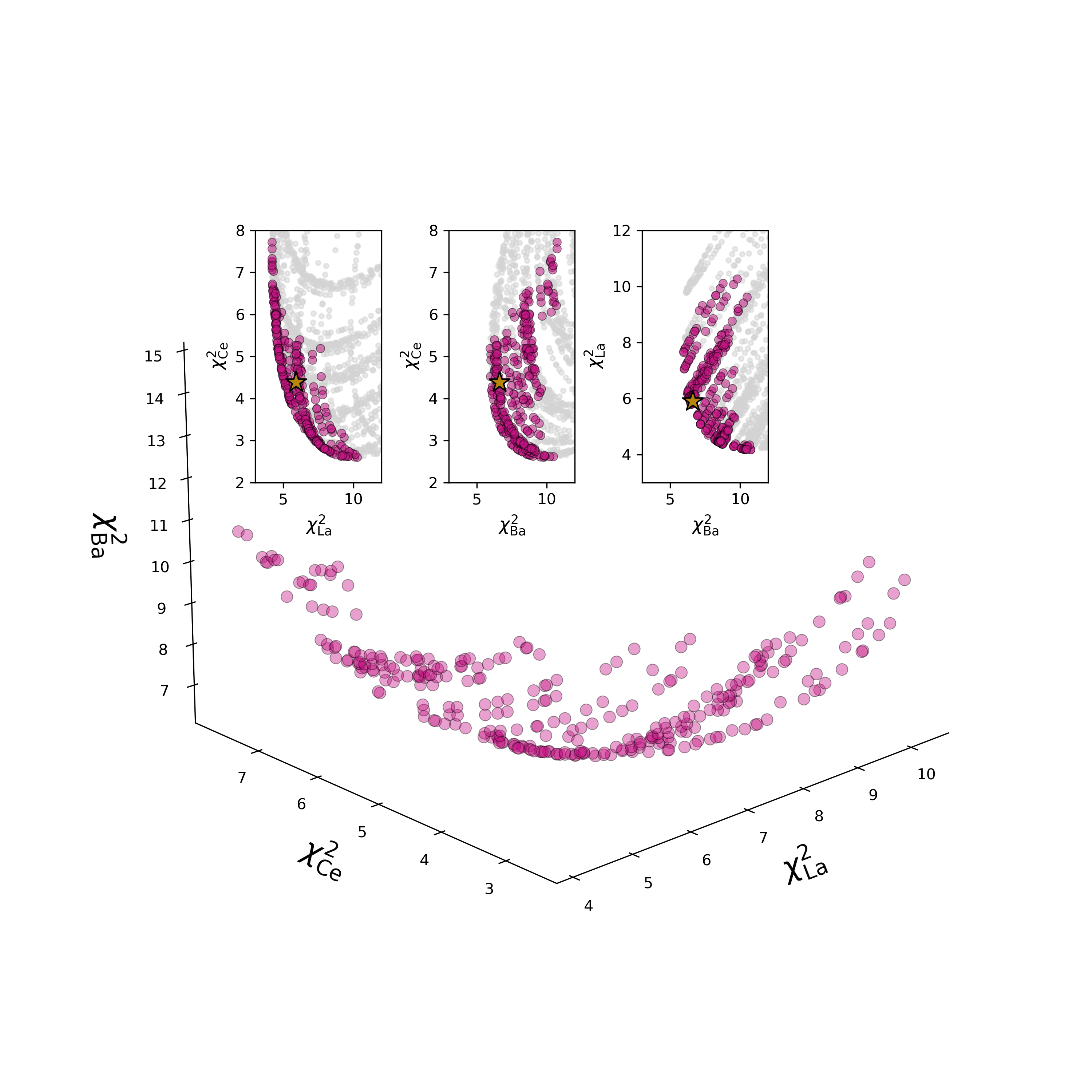}}
    \caption{Reduced $\chi^2$ values for [Sr/Fe], [Y/Fe] and [Zr/Fe] vs. [Fe/H] (left panel) and for [Ba/Fe], [La/Fe] and [Ce/Fe] (right panel) of the Pareto optimal solutions forming the three-dimensional Pareto surface. The insets show the corresponding two-dimensional projections of each pair of elements, with the gray points representing the full model distribution. The yellow star marks the model closest to the ideal point $(0,0,0)$ in this objective space. }
\label{fig: SrYZr_BaLaCe_chi2}
\end{center}
\end{figure*}

\begin{figure*}
    \includegraphics[width=1\textwidth]{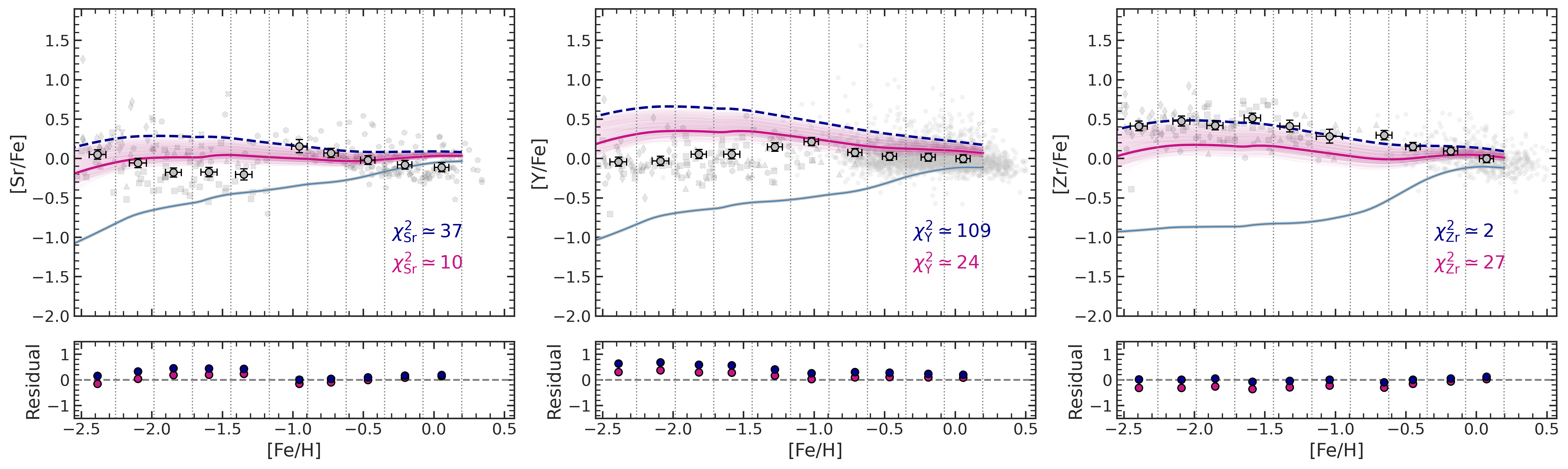}
     \caption{Comparison between observed [Sr/Fe], [Y/Fe] and [Zr/Fe] vs. [Fe/H] trends and the predictions of the Pareto optima models (magenta lines). The thicker magenta lines show the predicted [Sr/Fe], [Y/Fe] and [Zr/Fe] vs. [Fe/H] trends for the Pareto optimal model with the smallest distance to the ideal point $(0,0,0)$ in objective space. The dashed blue curve represents the best-fit model to Zr and its predictions for Sr and Y. The lower, light blue lines represent the results of the model in case there is only the s-process contribution. Grey circles with error bars show the average observational values in metallicity bins. Residual relative to the binned observations are displayed in the lower panels. Symbols for the observational data are the same as in Figure \ref{fig: best models EuFe vs FeH}.}%
 \label{fig: Sr_Y_Zr_FeH_pareto}%
\end{figure*}

The results of the three-dimensional optimization for Sr, Y and Zr is shown in Figure \ref{fig: SrYZr_BaLaCe_chi2} (left panel). In this case, the Pareto front corresponds to a surface in the three-dimensional objective space defined by the individual $\chi^2$ values of Sr, Y, and Zr. Each point on this surface represents a model for which no further improvement is possible in one element without simultaneously worsening the fit to at least one of the others. The Pareto front shows the trade-offs among the three elements. One can note that the distribution of models indicates that the $\chi^2$ for Sr and Zr can reach lower values than that of Y. Such a behaviour is consistent with the results discussed in Section \ref{sec: best-fit models}: the models that best reproduced Y are those that underestimate Sr and Zr. Therefore, in the multi-objective framework, the optimization reduces the discrepancies for Sr and Zr, at the expense of worsening the agreement with Y. In Table \ref{tab: model results}, we report the properties of the Pareto model closest to the \textbf{0} point in parameter space (the distributions of the parameters of all the Pareto-optimal models are reported in the left panel of Figure \ref{fig: SrYZr_pairplot BaLaCe_pairplot}). These solutions are close to those identified in the single-objective Eu analysis and in the two-dimensional Eu–Zr case. In particular, short delay times $\tau$ and progenitors in the lower end of the explored mass range are again favoured in order to reproduce the observed abundance trends. This is not unexpected: the observational data display a well-defined plateau at low metallicities, which can only be reproduced through a rapid injection of r-process material. As a consequence, no significant change in the preferred delay times is expected when moving from the Eu-only to the multi-element optimizations. The parameter that shows the largest variation across the different analyses is the effective Eu injection rate, $\alpha \times Y_{\rm Eu}$, which tends to increase, though it remains within the same order of magnitude as before. Aside from this adjustment, the models show no preference for a distinct region of the parameter space. The Pareto-optimal solutions are therefore primarily balancing the amount of r-process material injected, in order to shift the [Sr/Fe] and [Zr/Fe] trends toward higher values while retaining the level of agreement already achieved for [Y/Fe]. Such a trade-off is, however, not straightforward, as one can see from the Pareto optimal model results in Figure \ref{fig: Sr_Y_Zr_FeH_pareto}. 

In order to achieve a better agreement with the [Sr/Fe] and, in particular, with the [Zr/Fe] vs. [Fe/H] (magenta models in the upper part of the Pareto-optimal band), the fit to the observed [Y/Fe] vs. [Fe/H] is reduced. Indeed, the Pareto optimal model (with the shortest distance to the solution \textbf{0} in objective space) shows difficulties in reproducing the three trends at the same time. While it is capable to provide a satisfactory match to Sr, it fails to maintain the fit to Y, especially at low [Fe/H], and continues to predict a [Zr/Fe] trend which is too low, except for the two highest metallicity bins. For comparison, we also show the model that provides the best fit to the [Zr/Fe] vs. [Fe/H] relation, together with its corresponding predictions for Sr and Y (blue lines in Figure \ref{fig: Sr_Y_Zr_FeH_pareto}). As expected, this model overestimates the [Y/Fe] vs. [Fe/H] trend. It also tends to overestimate [Sr/Fe] at low metallicities, while providing a better fit at higher [Fe/H]. This inability of the models to fit Sr, Y and Zr at the same time may reflect differences between the solar r-process abundance distribution and that observed in low-metallicity r-process–enhanced stars (e.g., CS 22892–052, \citealp{Sneden2003}; HD 222926, \citealp{Roederer2018}), which typically exhibit higher Sr/Y and Zr/Y ratios than the solar pattern.

\subsubsection{Ba, La, Ce}
\label{sec: Ba, La, Ce}

We now extend the analysis to heavier elements, namely Ba, La and Ce. As discussed previously, also in this case these elements are not purely produced by the r-process. Nevertheless, they provide a statistically robust dataset for our multi-objective optimization study, thanks to the wealth of observational data available across a broad metallicity range. As for Sr, Y and Zr, we first study Ba, La and Ce simultaneously, independently of Eu. 

The Pareto front surface resulting from the three-dimensional optimization is shown in the right panel of Figure~\ref{fig: SrYZr_BaLaCe_chi2}. Compared to the case of Sr, Y, and Zr, the reduced $\chi^2$ values achieved here are significantly lower, and none of the elements shows a systematically better agreement than the others. The number of Pareto-optimal models is also smaller compared to the previous case ($\sim300$ Pareto-optimal models in the case of Ba, La and Ce, compared to $\sim 800$ in the case of Sr, Y and Zr). This reflects the reduced level of trade-offs among Ba, La and Ce, which can be indeed reproduced more consistently under the solar-scaled prescription and therefore fewer distinct solutions exist in the objective space, which causes a less populated Pareto surface.

The properties of the Pareto-optimal models, in terms of input parameter values, are broadly consistent with those obtained in the single-objective (Eu-only) and earlier multi-objective optimizations (see right panel in Figure \ref{fig: SrYZr_pairplot BaLaCe_pairplot}). As expected, short delay times are favoured, and most solutions correspond to progenitors in the lower end of the explored mass range. No models with progenitor masses above $70\ \rm M_\odot$ are selected, a slightly stricter cutoff than in the previous case (where the limit extended to $80\ \rm M_\odot$). Among the Pareto-optimal models, the solution closest to the \textbf{0} point in the objective space, corresponds to a lower progenitor mass range, combined with a slightly lower Eu injection rate (see Table \ref{tab: model results}). The two parameters are to some extent degenerate: due to the IMF weighting, lower-mass progenitors must be compensated by lower r-process injection rate. Nevertheless, the overall distribution still shows a clear preference for lower progenitor masses (see the $M_{\rm mid}$ vs. $\alpha \times Y_{\rm Eu}$ panel of Figure \ref{fig: SrYZr_pairplot BaLaCe_pairplot}).

\begin{figure*}
    \includegraphics[width=1\textwidth]{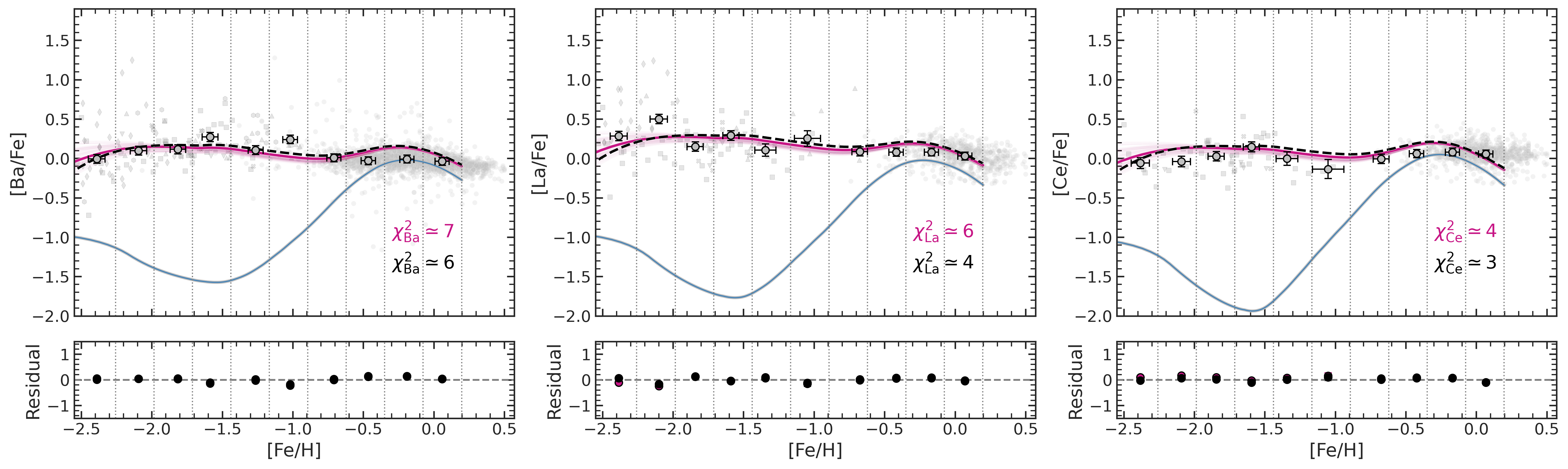}
     \caption{Comparison between observed [Ba/Fe], [La/Fe] and [Ce/Fe] vs. [Fe/H] trends and the predictions of the Pareto optima models (magenta lines). The thicker magenta lines show the predicted [Ba/Fe], [La/Fe] and [Ce/Fe] vs. [Fe/H] trends for the Pareto optimal model with the smallest distance to the ideal point $(0,0,0)$ in objective space. The dashed black curves represents the best-fit model to the specific element. The lower, light blue lines represent the results of the model in case there is only the s-process contribution. Grey circles with error bars show the average observational values in metallicity bins. Residual relative to the binned observations are displayed in the lower panels. Symbols for the observational data are the same as in Figure \ref{fig: best models EuFe vs FeH}.}%
 \label{fig: Ba_La_Ce_FeH_pareto}%
\end{figure*}

The reduced $\chi^2$ values of the best Pareto-optimal model for Ba, La, and Ce are comparable to those obtained in the single-element analyses, indicating that all three elements can be simultaneously reproduced by the same model. This consistency is clearly reflected in the predicted [X/Fe] vs. [Fe/H] trends, shown in Figure \ref{fig: Ba_La_Ce_FeH_pareto}. The overall abundance ratios trends are extremely similar, with only minor deviations visible at very low metallicity. Moreover, the results of the single-element best models closely match not only the Pareto-optimal solution closest to the \textbf{0} one, but also the full set of Pareto-optimal models. Such an agreement highlights the robustness of the r-process pattern beyond the second peak, in contrast to the lighter r-process elements.

\subsubsection{Global comparison}

\begin{figure*}
\begin{center}
    \subfloat{\includegraphics[width=0.4\textwidth]{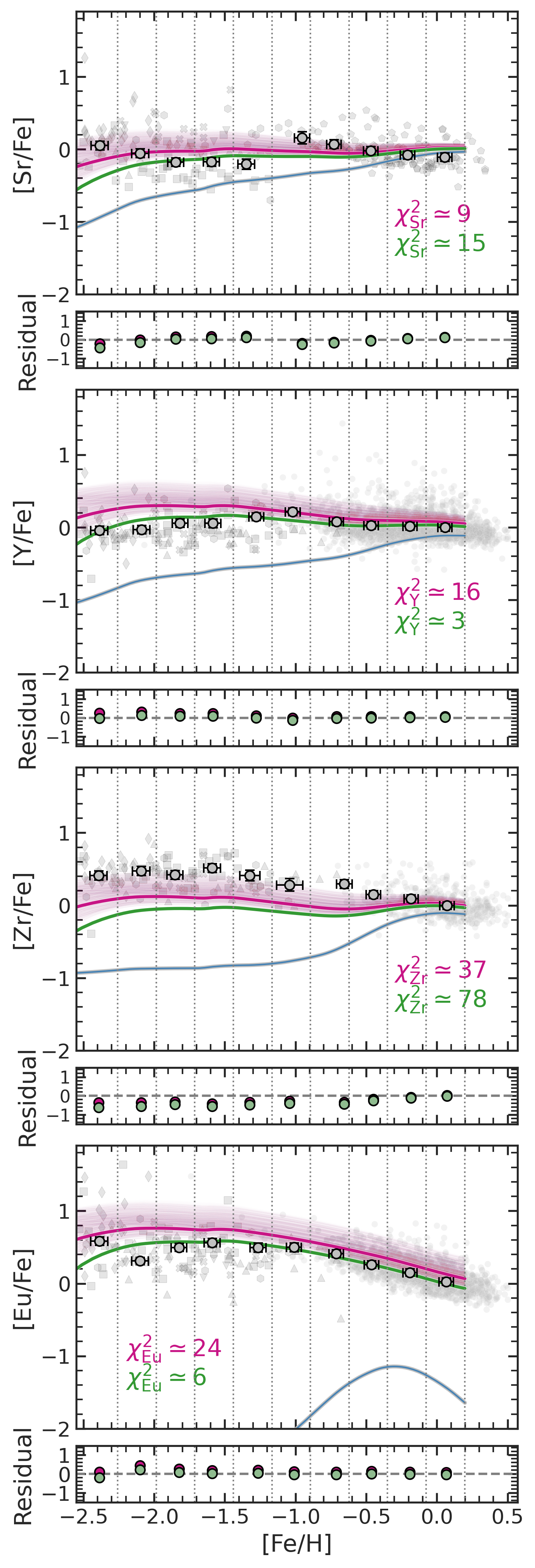}}
    \hspace{1cm}
    \subfloat{\includegraphics[width=0.4\textwidth]{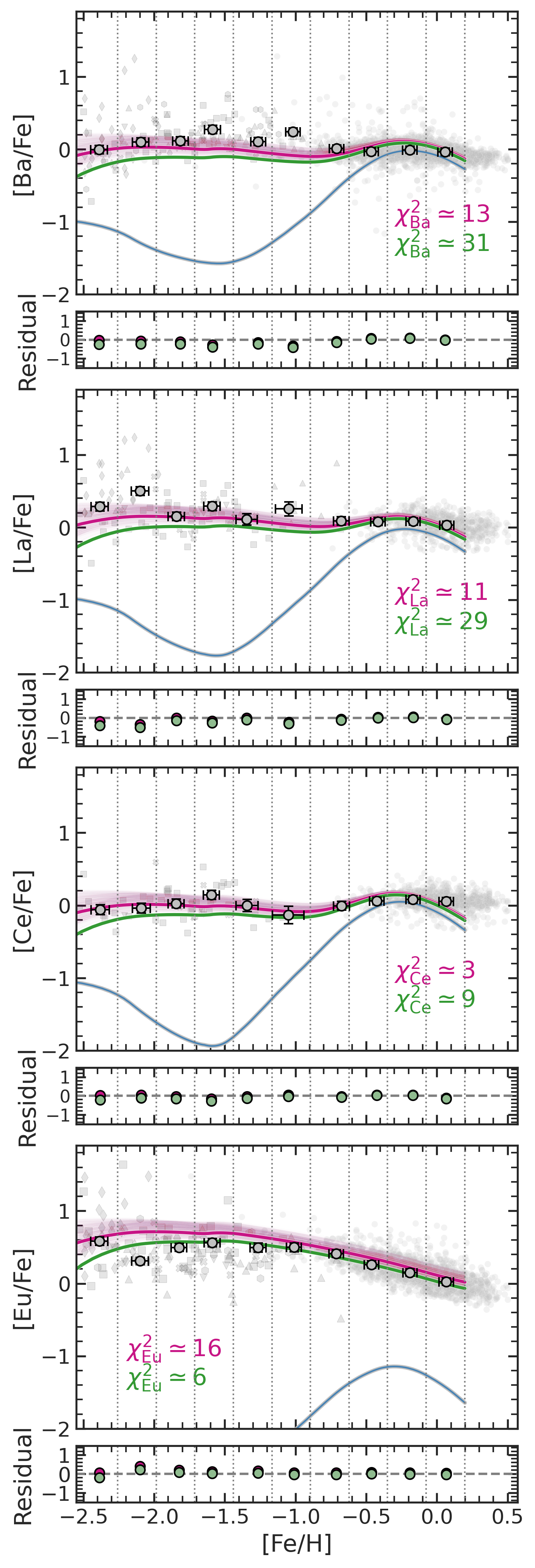}}
     \caption{Comparison between observed [X/Fe] vs. [Fe/H] trends and the predictions of the Pareto optima models (magenta lines) for the Sr-Y-Zr-Eu group (left column) and the Ba-La-Ce-Eu group (right panel). The thicker magenta lines show the predicted [Eu/Fe], [Sr/Fe], [Y/Fe] and [Zr/Fe] vs. [Fe/H] trends for the Pareto optimal model with the smallest distance to the ideal point $(0,0,0,0)$ in objective space (see Table \ref{tab: model results} for parameter values). The solid green curve shows the best-fit model to Eu and its corresponding predictions for the other elements. Grey circles with error bars show the average observational values in metallicity bins. Residuals relative to the binned observations are displayed in the lower panels. Symbols for the observational data are the same as in Figure \ref{fig: best models EuFe vs FeH}.}%
 \label{fig: Eu_Sr_Y_Zr_FeH_pareto}%
\end{center}
\end{figure*}

In the previous sections we focused on the elements groups separately, first the first-peak (Sr, Y, Zr) elements and then the second-peak (Ba, La, Ce), independently on Eu. In this section, we perform a global comparison, by including Eu together with each of the two groups in order to test the overall validity of the solar-scaled r-process prescription and to understand whether any limitation is confined to the lighter elements or may also extend to the heavier ones.

As reported in Table \ref{tab: model results} (see also left panel of Figure \ref{fig: EuSrYZr_pairplot EuBaLaCe_pairplot}), the properties of the best Pareto-optimal solution for the four-dimensional case represented by the combined set of Sr, Y, Zr and Eu are similar to the ones obtained in the three-dimensional case discussed in Section \ref{sec: Sr, Y, Zr}. The parameters resemble those obtained in the previous analysis, with short delay times and lower progenitor mass ranges being favoured. The effective Eu injection rate, $\alpha \times Y_{\rm Eu}$, tends to be lower than in the simpler cases, reflecting the fact that, among the lighter elements, only Y is well reproduced by the best Eu-only model, whereas Sr and Zr are systematically underestimated. The Pareto optimization therefore attempts to increase the effective Eu injection rate to improve the fit to the [Sr/Fe] and [Zr/Fe] vs. [Fe/H] relations. This increase, however, is limited since pushing the injection rate to too high values would worsen the agreement with both Eu and Y. The resulting Pareto-optimal solutions thus have to balance the improvement in Sr and Zr without compromising the already obtained agreement for Eu and Y.

Such a behaviour is visible in the left column of Figure \ref{fig: Eu_Sr_Y_Zr_FeH_pareto}, where both the Pareto-optimal models and the best Eu-only model predictions are shown. The trade-offs are evident, as the model that reproduce Eu and Y (green lines) underestimates Sr and especially Zr, while the best Pareto-optimal model improves the agreement with Sr and Zr but at the expense of Eu and Y. Notably, however, the worsening of the fit to the [Eu/Fe] and [Y/Fe] is confined to the low-metallicity regime ($\mathrm{[Fe/H] \lesssim -1.1\ dex}$), whereas the higher-metallicity range remains well reproduced. This model achieves a good agreement with the [Sr/Fe] vs. [Fe/H] trend, but the [Zr/Fe] one still remains underestimated over nearly the entire metallicity range, reflecting the initial condition that Sr was less underestimated than Zr in the Eu-only solutions.

The situation is notably different when Eu is combined with Ba, La and Ce. The Pareto-optimal models occupy a parameter space consistent with that found in the previous cases (see right panel of Figure \ref{fig: EuSrYZr_pairplot EuBaLaCe_pairplot}): short delay times and lower progenitor masses are favoured, and no solutions extend beyond $80 \rm\ M_\odot$. The effective Eu injection rate, $\alpha \times Y_{\rm Eu}$, is again the only parameter that changes, with a lower value achieved by the best Pareto model, as reported in Table \ref{tab: model results}. 

Within this parameter space, the Pareto-optimal solutions reveal no strong conflicts among the four elements, that can be reproduced simultaneously by many of the Pareto models. This is shown in the right panel of Figure \ref{fig: EuSrYZr_pairplot EuBaLaCe_pairplot}, where one can see that the reduced $\chi^2$ values for the best Pareto solutions are improved with respect to those obtained in the Eu-only optimization. As expected, the agreement with Eu is partially reduced in this case, reaching, however, a better agreement to that found in the first-peak analysis. However, the loss is not drastic, and the overall outcome represents a satisfactory compromise that allows all four elements to be consistently reproduced.

Overall, compared to the Sr, Y, Zr and Eu case, the Pareto-optimal models obtained in the four-dimensional case with Ba, La, Ce and Eu achieve systematically lower values of the distance $D$ to the ideal point $(0,0,0,0)$ in objective space. The maximum distance is $D \simeq 50$, whereas for lighter elements it reaches values as high as $D \simeq 200$. Also, the distance of the best Pareto model is around half of that obtained for the Sr, Y, Zr, Eu group. In addition, the number of Pareto-optimal models ($\sim 800$) is smaller than in the first-peak case ($\sim 1200$), implying that the trade-offs among Ba, La, Ce and Eu are less severe and that the solution space is more tightly constrained (as also discussed in section \ref{sec: Ba, La, Ce}). 

\section{Different r-process patterns}
\label{sec: different scaling relations}

\begin{figure}
    \includegraphics[width=1\columnwidth]{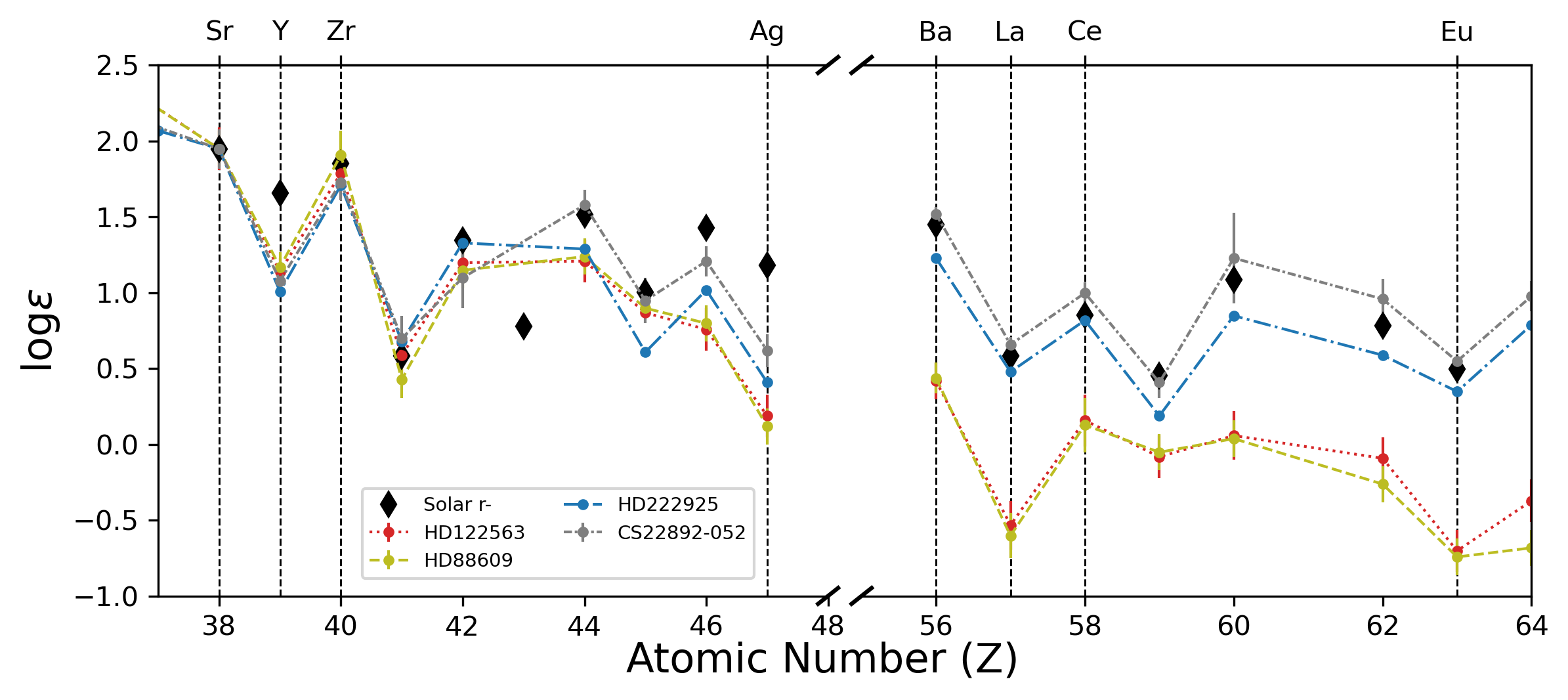}
     \caption{Abundance patterns of four r-process stars compared to the solar r-process distribution, normalised to the solar $\rm log\epsilon_{Sr}$. The stars include HD 122563 and HD 88609 (r- poor stars) and HD 222925 and CS22892-052 (r- rich stars).}%
 \label{fig: stars}%
\end{figure}

The concept of r-process universality arises from the similarity between the relative abundance patterns of lanthanides and third-peak r-process elements observed in many r-process rich stars and those in the Sun. This universality suggests that the r-process operates in a robust manner across different sites and epochs of Galactic evolution (e.g., \citealp{Westin2000, Frebel2018, Sakari2018, Cowan2021, Roederer2022}). Observations, however, suggest that the universality of the r-process may extend to the second r-process peak, but not to the lighter neutron-capture elements (e.g., \citealp{Qian2008, Mashonkina2010, Hansen2012, Hansen2014, Aoki2017, Spite2018}). The results of this work, showing a closer agreement with the heavier neutron-capture elements than with the lighter ones when adopting solar r-process scaling with Eu, are consistent with the observed difficulty of extending r-process robustness toward the lighter elements. As discussed in section \ref{sec: best-fit models}, the scaling relation itself may depend on metallicity, favouring higher Sr and Zr production as one moves towards lower metallicity. Moreover, the abundance patterns of metal-poor, r-poor stars (e.g. HD 122563 and HD 88609, known also as the Honda stars or r-limited stars) display relative enhancements in Sr, Y, and Zr, when normalized to Eu. Such an enhancement may reflect an additional nucleosynthesis contribution, different from the main r-process, which indeed operates predominantly early during Galactic evolution (e.g., \citealp{Wasserburg1996, Travaglio2004, Montes2007, CescuttiChiappini2010, Psaltis2024}). Contrary to r-poor stars, r-rich stars (e.g. HD 222925 and CS 22892-052) exhibit a solar-like abundance pattern, except that for Sr, Y and Zr, for which both classes of stars display ratios of Sr/Y and of Zr/Y consistently exceeding the solar values (see Figure \ref{fig: stars}; see also \citealp{Racca2025}).

\begin{figure*}
    \includegraphics[width=1\textwidth]{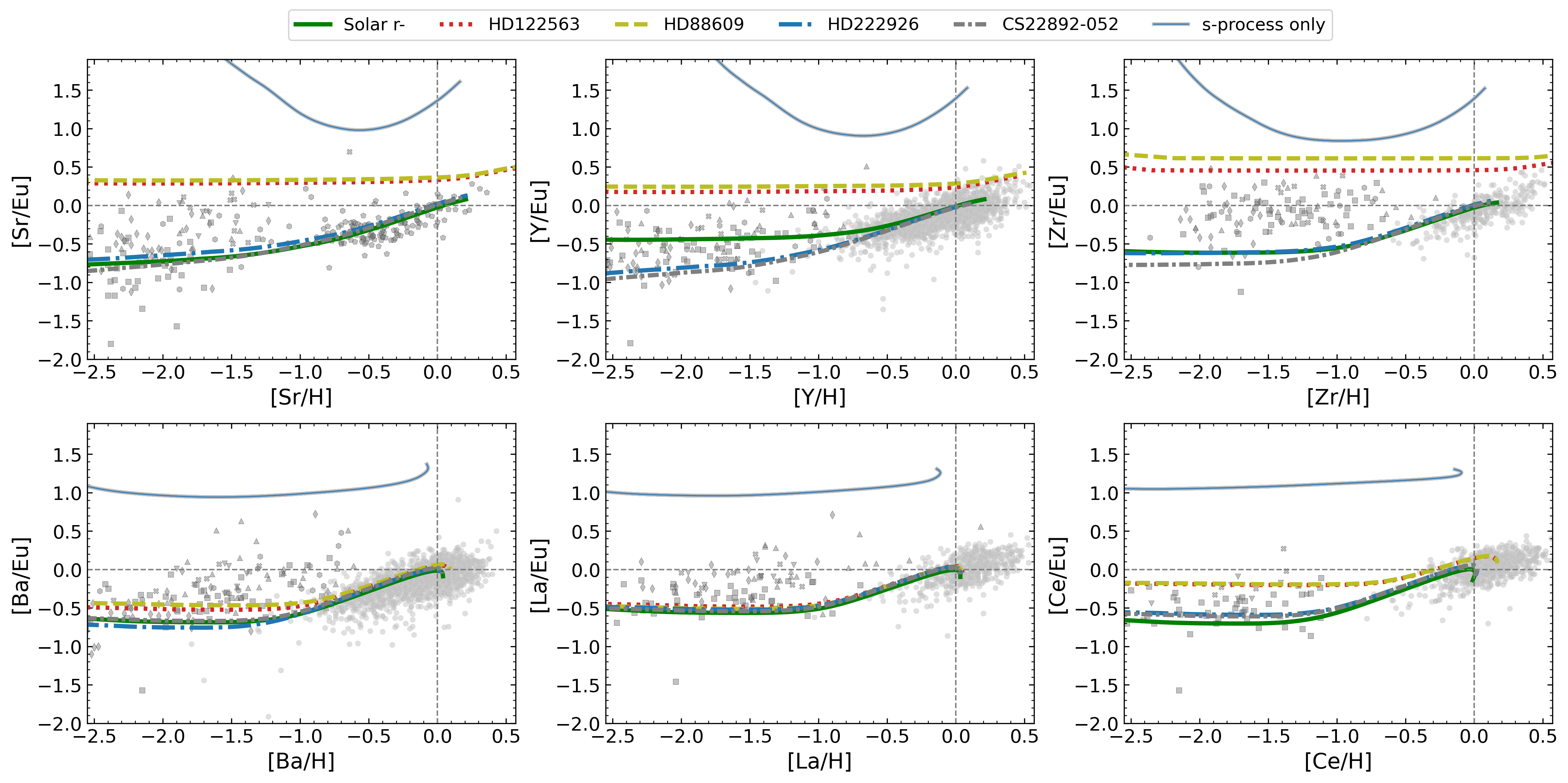}
     \caption{Observed and predicted abundance ratios [X/Eu] vs. [X/H] for the first- and second-peak neutron-capture elements (Sr, Y, Zr, Ba, La, Ce). Grey symbols show observational data with the same symbol types as in previous figures, while colored lines represent model predictions using different scaling relations: the solar r-/s-process pattern (solid green/gray), HD 122563 (red dotted), HD 88609 (yellow dashed), HD 222925 (blue dash-dotted), and CS 22892-052 (grey dash-dotted). Symbols for the observational data are the same as in Figure \ref{fig: best models EuFe vs FeH}.}%
 \label{fig: ncEu_1st_2nd_ncH}%
\end{figure*}

Figure \ref{fig: ncEu_1st_2nd_ncH} shows the Eu-only best-fit model from section \ref{sec: Eu optimization}, now rescaled using abundance patterns from the four stars HD 122563, HD 88609, HD 222925 and CS 22892-052. In order to focus only on the neutron-capture elements and their ratios with Eu, we removed the dependence on Fe and plot the observed and predicted [X/Eu] vs. [X/H] trends. In this way we can also see, in both the models and the observational data, a plateau in the [X/Eu] vs. [X/H] ratio at intermediate to low [X/H], which indicates a similar production timescale for the two elements. This is followed by an increase at higher [X/H], corresponding to the onset of the s-process contribution from AGB stars to the production of the element X (see also \citealp{Hansen2012, Lombardo2025}). The different lines in the figure correspond to the various scaling relations. When the relative elements production with r-poor stars, the lighter element trends are systematically overestimated across the entire [X/H] range. Cerium is also overproduced, although only within the plateau region. In contrast, scaling with r-rich stars yields results similar to those obtained with the solar r-process pattern, except that for Y, as expected.

The four stars are sufficiently metal poor ($\rm [Fe/H] \lesssim -1.46\ dex$) to be considered r-process dominated and, for HD 88609 and CS 22892-052 (an r-poor and an r-rich star, respectively), the metallicities are low enough ($\rm [Fe/H] \lesssim -3\ dex$) that their abundance patterns could reflect enrichment by a single, or a few, r-process events, consistent with the stochastic nature of chemical enrichment towards earlier times. Therefore, the event that enriched an r-poor star may not be the same as that which enriched an r-rich star, so that at least two different types of events would be needed to explain both classes of stars. As proposed by \citet{QianWasserburg2000}, one class of events would be capable of producing heavy r-process elements (known as the H-component), yielding an abundance pattern which resembles that of r-rich stars, while another would account for the lighter elements enhanced patterns characteristic of r-poor stars (known as the L-component). Previous work from \citet{Hansen2014} (see also \citealp{Li2013}) have already shown that the abundance patterns of many low-metallicity stars can be reproduced by these two components, with the abundances of elements $\rm Z \geq 56$ primarily produced by the H-components and the abundance of elements $\rm 38 \leq Z \leq 47$ created by a combination of L- and H-components (see also \citealp{Qian2007}).

In a Galactic chemical evolution framework, such a scenario can be tested in a site-independent manner by adopting multiple scaling relations that may depend on metallicity and/or the evolutionary stage of the Galaxy. In this context, the component responsible for the production of the lighter neutron-capture elements may need to be constrained directly within the chemical evolution model, or alternatively inferred from simulated abundance patterns that reflect different astrophysical conditions, such as those recently presented by \citet{Kuske2025}. Both approaches will be explored in future work.

\section{Conclusions}
\label{sec: conclusions}

In this work, we compute $\sim 1.5 \times 10^5$ Galactic chemical evolution models, systematically varying the parameters associated with r-process nucleosynthesis: the Eu yield per event, the event rate, the delay time before enrichment, and the progenitor star mass range. Rather than assuming a specific physical site, we adopt a site-independent, parametric approach to explore a wide range of astrophysical conditions under which the r-process enrichment could occur. We compare the resulting [X/Fe] vs. [Fe/H] Galactic chemical evolution trends with high-resolution observations of multiple neutron-capture elements. The yields of these elements are derived from Eu by assuming solar r-process scaling relations, thereby testing whether the robustness of the r-process extends to lighter neutron-capture elements. To account for the trade-offs among different observational constraints, we frame the comparison as a multi-objective optimization problem and identify the set of Pareto-optimal solutions across various element combinations. 
\\
\\
Our main results are:
\begin{itemize}
    \item The best models for reproducing the observed [Eu/Fe] vs. [Fe/H] trend favor low mass progenitor stars ($M_{\rm mid}=22.5\ \rm M_\odot$), quick delay times for the pollution ($\tau = 30\ \rm Myr$) and an effective Eu injection rate of $\sim 2\times 10^{-7}\ \rm M_\odot$ per event.
    \item A degeneracy exists between the progenitor mass range and the effective Eu injection rate, as higher-mass progenitors require larger yields owing to their lower frequency. Progenitors above $\sim 80\ \rm M_\odot$ cannot account for the observed Eu production, even when assuming high effective yields, because they are too rare. This does not exclude the possibility of having rare r-process events from very high-mass progenitors, but it implies that they can not represent the dominant r-process site over cosmic time.
    \item The best-Eu models, once scaled according to the solar r-process abundance pattern, show inconsistencies into reproducing lighter elements at intermediate to low metallicities, with Sr, Zr, Ba and La being underestimated, and Zr being the more problematic.
    \item In the two-dimensional Eu–Zr case, the Pareto-optimal solutions struggle to simultaneously reproduce both elements: [Zr/Fe] is matched only by those models that systematically overestimate [Eu/Fe], pointing to no robustness of the r-process towards lighter elements.
    \item Independently of Eu, the Pareto-optimal models show greater difficulty in identifying a common set of nucleosynthetic parameters capable of reproducing the lighter elements (Sr, Y, Zr) compared to the heavier ones (Ba, La, Ce). The less coherence obtained for the lighter elements is due to the different Sr/Y and Zr/Y observed in the Sun with respect to that observed in the other r-process dominated stars.
    \item When combined with Eu, Pareto-optimal models reveal a stronger conflict among Eu, Sr, Y, Zr than with Eu, Ba, La, Ce, reaching systematically higher values of the distance $D$ to the ideal point (0,0,0,0) in objective space, confirming that the assumption of a solar-scaling relation becomes progressively less valid toward the lighter neutron-capture elements.
\end{itemize}

In conclusion, we expect at least two distinct classes of scaling relations: one associated with events that reproduce the main r-process component, yielding an abundance pattern comparable to the solar r-process (or that of r-rich stars; H-component), and another representative of a weak r-process, characterized by enhanced production of Sr, Y, and Zr relative to the lanthanides (similar to the one observed in r-poor stars; L-component). Nevertheless, an L-component directly inferred from the abundance patterns of r-poor stars such as HD 122563 and HD 88609 leads to an excessively strong enhancement in the resulting abundance ratios. Instead, such a component may be constrained within the chemical evolution model itself or derived from simulated abundance patterns.

\begin{acknowledgements}
We thank Jan Kuske for the useful discussions. This work was supported by the Deutsche Forschungsgemeinschaft
(DFG, German Research Foundation) – Project-ID 279384907 – SFB 1245. We acknowledge support by the National Science Foundation under Grant No. OISE-1927130 (IReNA). F.M. thanks the support by the National Science Foundation under grant PHY-2209429.

\end{acknowledgements}

%
%

\bibliographystyle{aa} 
\bibliography{Bibliography} 


\begin{appendix}

\section{Corner plots for the Pareto-optimal models}

\begin{figure}
    \includegraphics[width=1\columnwidth]{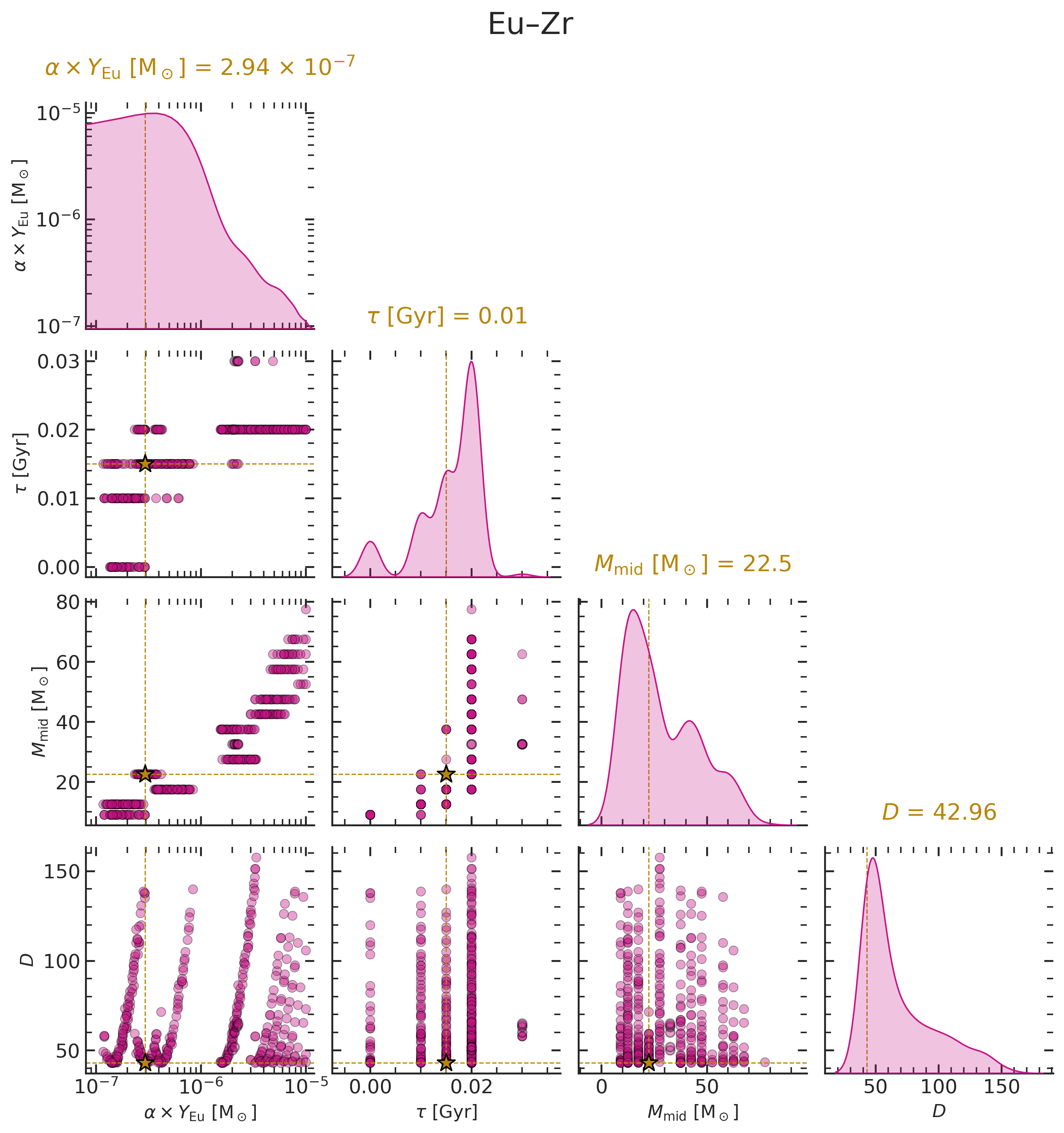}
     \caption{Corner plot showing the marginal distributions (diagonal panels) and pairwise correlations (off-diagonal panels) of the parameters $(Y_{\rm Eu,\rm r},\ \alpha,\ [M_{\rm l}, M_{\rm u}],\ \tau)$ for the (Eu-Zr) Pareto models. Dashed yellow lines mark the parameter values of the model closest to the ideal point $(0,0)$ in the objective space, with the corresponding numerical values reported above each marginal distribution, and the yellow star marks its position in each correlation plot.}%
 \label{fig: pairplot Pareto models Eu-Zr}%
\end{figure}

\begin{figure*}
    \subfloat{\includegraphics[width=0.45\textwidth]{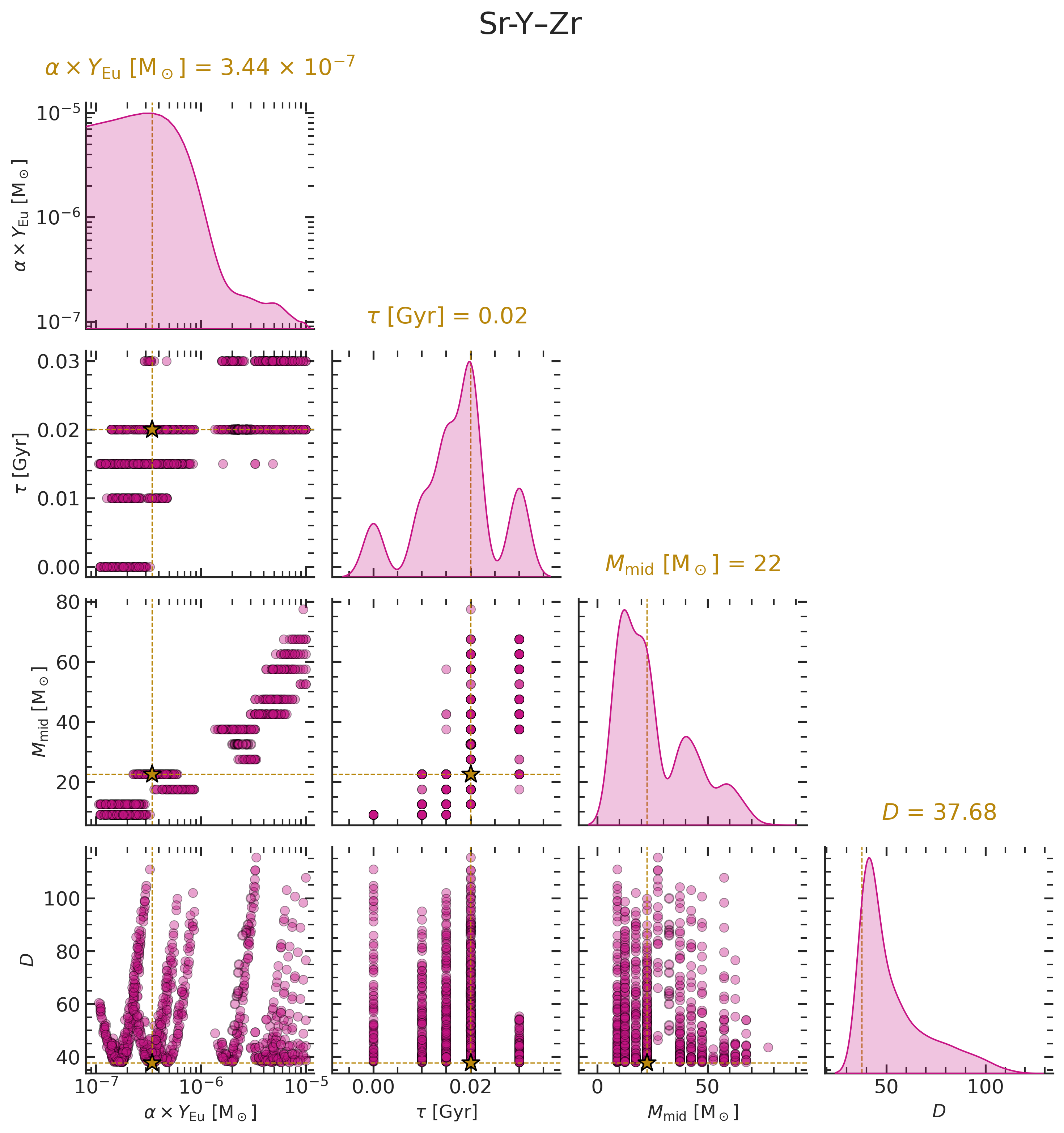}}
    \hfill{}
    \subfloat{\includegraphics[width=0.45\textwidth]{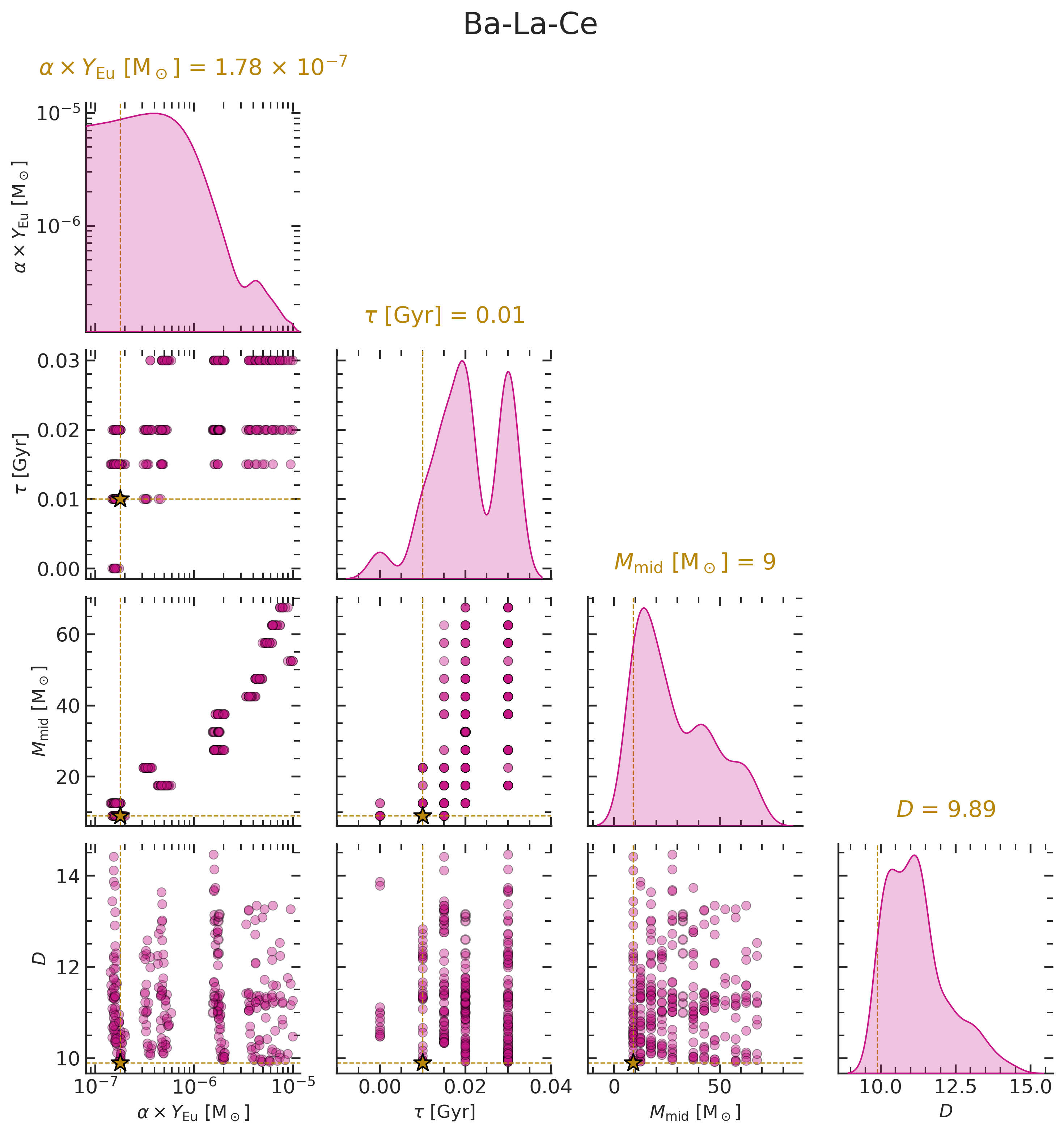}}
    \caption{Same as Figure \ref{fig: pairplot Pareto models Eu-Zr} but for Sr, Y, Zr (left panel) and for Ba, La, Ce (right panel).}
\label{fig: SrYZr_pairplot BaLaCe_pairplot}
\end{figure*}

\begin{figure*}
    \subfloat{\includegraphics[width=0.45\textwidth]{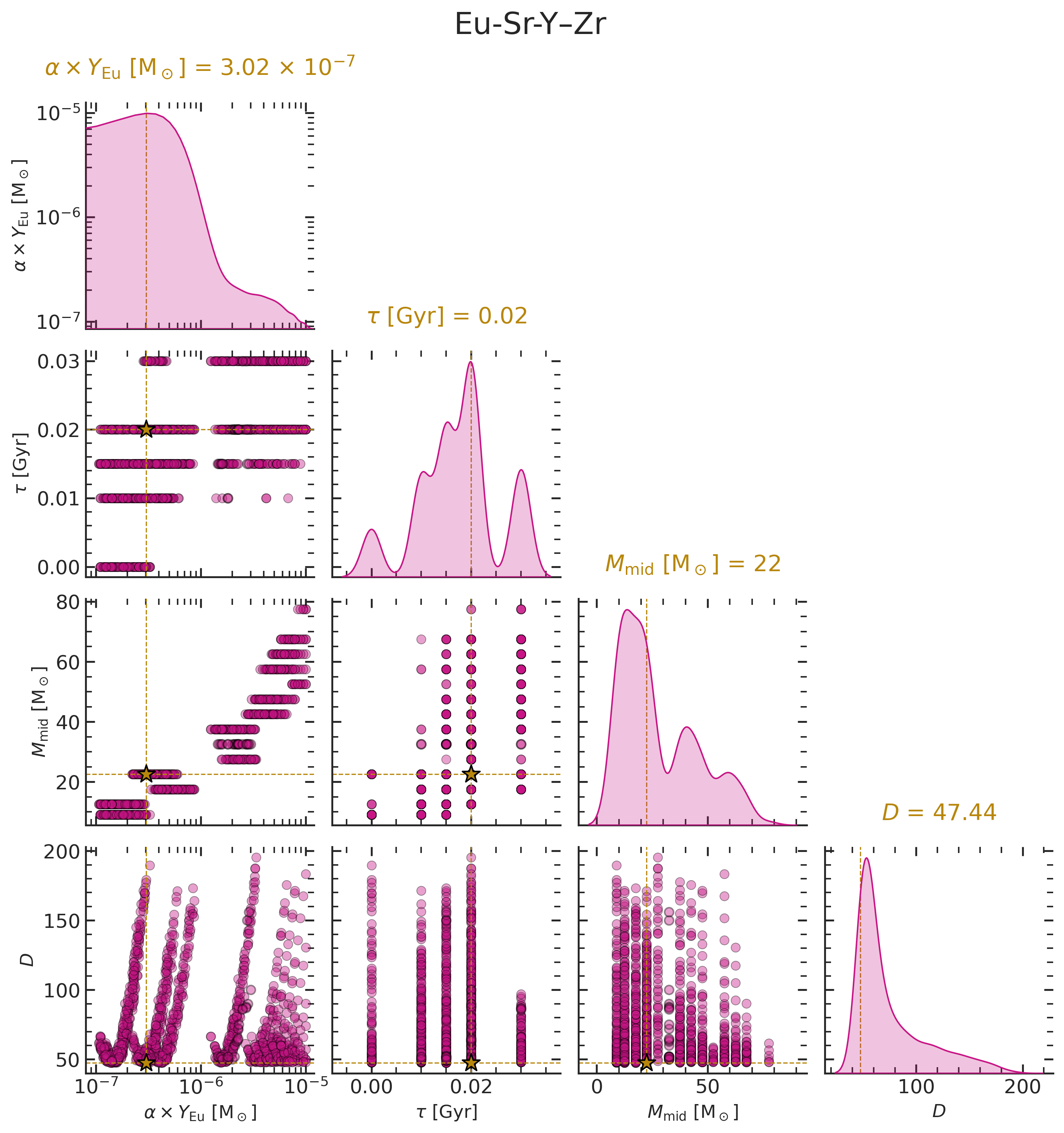}}
    \hfill{}
    \subfloat{\includegraphics[width=0.45\textwidth]{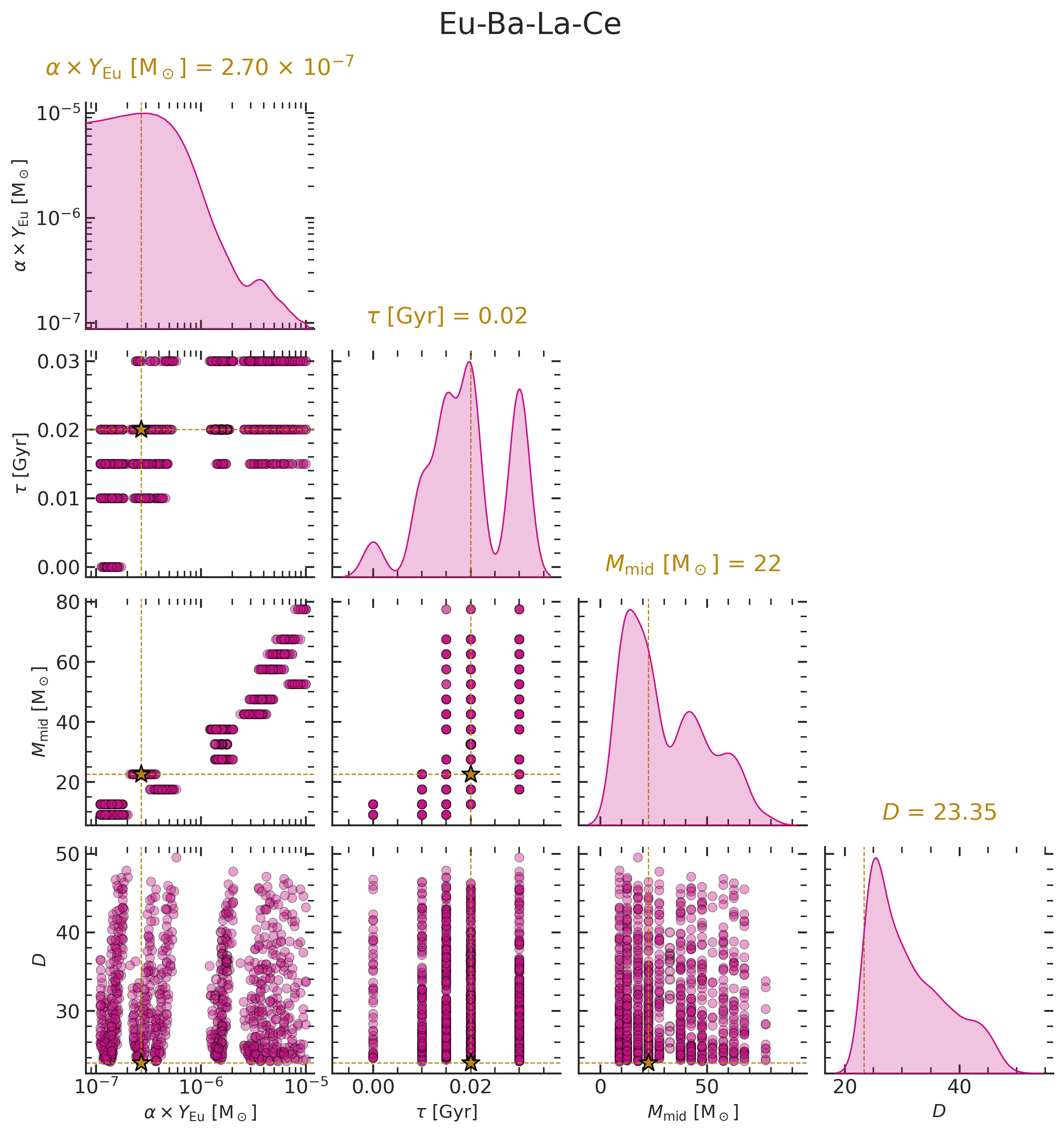}}
    \caption{Same as Figure \ref{fig: pairplot Pareto models Eu-Zr} but for Sr, Y, Zr, Eu (left panel) and for Ba, La, Ce, Eu (right panel).}
\label{fig: EuSrYZr_pairplot EuBaLaCe_pairplot}
\end{figure*}

In this appendix we provide the complete set of corner plots discussed in the main text. Each panel shows the marginal distributions (diagonal) and pairwise correlations (off-diagonal) of the model parameters used in the multi-objective optimization for the different combinations of elements considered (Fig. \ref{fig: pairplot Pareto models Eu-Zr} to \ref{fig: EuSrYZr_pairplot EuBaLaCe_pairplot}).  

\section{Other neutron-capture elements} 

\begin{figure*}
    \includegraphics[width=1\textwidth]{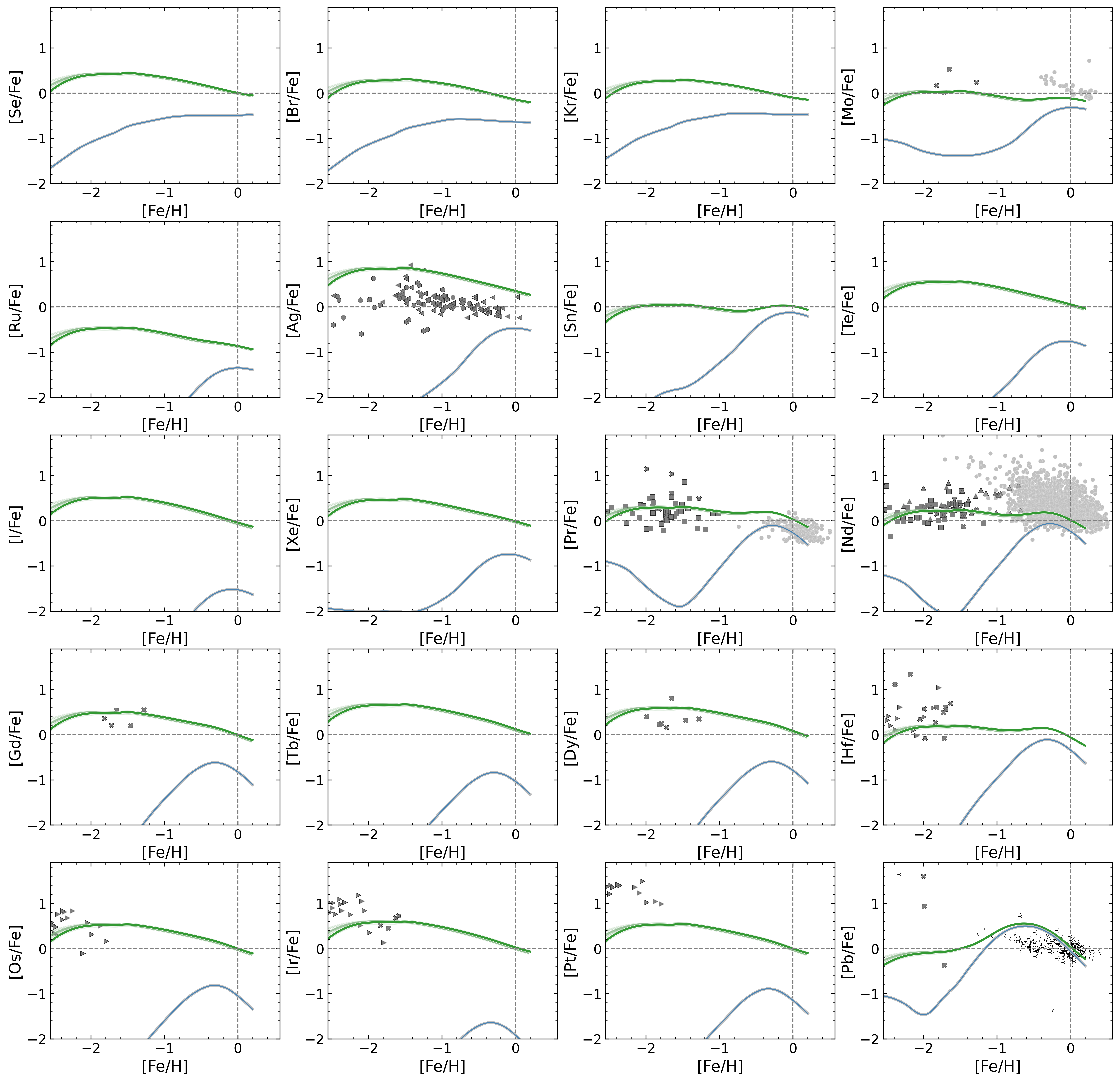}
     \caption{Comparison between observed [X/Fe] vs. [Fe/H] trends and the predictions of the best Eu models for the other neutron-capture elements included in the model. Symbols for the observational data are the same as in Figure \ref{fig: best models EuFe vs FeH}, with the addition of observations from \citet{Wu2015} for Ag (grey left-triangles), \citet{Alencastro2025} for Hf, Os, Ir and Pt (grey right-triangles) and from \citet{Contursi2024} for Pb (grey small right-triangles).}%
 \label{fig: other elements}%
\end{figure*}

As discussed in Section \ref{sec: Light and heavy neutron-capture elements}, the ideal Pareto-front models would include comparisons between Eu and elements belonging to the first r-process peak (Se, Br, Kr), the second peak (Te, I, Xe), the lanthanides (Gd, Tb, Dy), and the third peak (Os, Ir, Pt). However, the number of available observations for these elements across the relevant metallicity range is insufficient to enable a statistically meaningful comparison with the chemical evolution model. Therefore, here we show only predictions for these and other elements (see Figure \ref{fig: other elements}), which can be compared with the limited observational data available in the literature. In particular, we select data from the works discussed in Section \ref{sec: observational data}, together with some few other high resolution ($R \geq 40 000$) studies. We exclude all upper limits and include only CEMP-no stars. Although few other stars exist of the selected elements, their number remains insufficient to reveal any clear trend.

For many of the elements, data are not enough in number to allow us to draw strong conclusion about the quality of the models. It is evident, however, the overproduction of the [Ag/Fe] trend, across the entire range of metallicity. This is the effect of the adopted solar scaling, since, as shown in Figure \ref{fig: stars}, the solar abundance of Ag is higher than that observed in the other r-process dominated stars, again suggesting a reduced robustness of the r-process toward the lighter neutron-capture elements.

Regarding the third r-process peak, the model reproduces the abundances of Os and Ir reasonably well but overestimates that of Pt. Data are however too few to derive strong conclusion, highlighting the urgent need for additional observations. Pb is also an interesting case: the strong s-process contribution from AGB in the metallicity range $\mathrm{-1 \lesssim [Fe/H] \lesssim 0}$ dominates the predicted trend, causing an overestimation relative to observations. A detailed discussion of this effect is provided in \citet{Contursi2024}, where the same s-process chemical evolution models adopted this work are tested. We refer the reader to that work for more details.

\end{appendix}

\end{document}